\documentclass[preprint]{aastex} 
\newcommand{\about}     {\hbox{$\sim$}}
\newcommand{\x}         {\hbox{$\times$}}

\newcommand{\uM}        {\emph{u}}
\newcommand{\gM}        {\emph{g}}
\newcommand{\rM}        {\emph{r}}
\newcommand{\iM}        {\emph{i}}
\newcommand{\zM}        {\emph{z}}
\newcommand{\pt}[2]{$#1\times 10^{#2}$}
\newcommand{\si}[1]{\ensuremath{_{\textrm{\scriptsize{#1}}}}}

\begin{document}


\title{Double-Peaked Low-Ionization Emission Lines in Active Galactic Nuclei}

\author{
Iskra V. Strateva\altaffilmark{1}, 
Michael A. Strauss\altaffilmark{1},
Lei Hao\altaffilmark{1}, 
David J. Schlegel\altaffilmark{1},
Pat B. Hall\altaffilmark{1,2}, 
James E. Gunn\altaffilmark{1},
Li-Xin Li\altaffilmark{3}, 
\v{Z}eljko Ivezi\'{c}\altaffilmark{1},
Gordon T. Richards\altaffilmark{1}, 
Nadia L. Zakamska\altaffilmark{1},  
Wolfgang Voges\altaffilmark{4}, 
Scott F. Anderson\altaffilmark{5}, 
Robert H. Lupton\altaffilmark{1},
Donald P. Schneider\altaffilmark{6},
Jon Brinkmann\altaffilmark{7},
Robert C. Nichol\altaffilmark{8}
}

\altaffiltext{1}{Princeton University Observatory, Princeton, NJ
08544}
\altaffiltext{2}{Departamento de Astronom\'{\i}a y Astrof\'{\i}sica,
Facultad de F\'{\i}sica, Pontificia Universidad Cat\'{o}lica de Chile,
Casilla 306, Santiago 22, Chile}
\altaffiltext{3}{Harvard-Smithsonian Center for Astrophysics, 60 
Garden Street, Cambridge, MA 02138}
\altaffiltext{4}{Max-Planck-Institut f$\ddot{\textrm{u}}$r extraterrestrische Physik
Giessenbachstr.1, D-85748 Garching, Germany}
\altaffiltext{5}{Astronomy Department, University of Washington, Box
  351580, Seattle, WA 98195-1580}
\altaffiltext{6}{Department of Astronomy and Astrophysics, 504 Davey Laboratory, 
University Park, PA 16802}
\altaffiltext{7}{Apache Point Observatory, P.O. Box 59, Sunspot, NM 88349-0059}
\altaffiltext{8}{Department of Physics, Carnegie Mellon University, Pittsburgh, PA 15213}

\begin{abstract}
  We present a new sample of 116 double-peaked Balmer line Active
  Galactic Nuclei (AGNs) selected from the Sloan Digital Sky Survey.
  Double-peaked emission lines are believed to originate in the
  accretion disks of AGN, a few hundred gravitational radii
  ($R\si{G}$) from the supermassive black hole. We investigate the
  properties of the candidate disk emitters with respect to the full
  sample of AGN over the same redshifts, focusing on optical, radio
  and X-ray flux, broad line shapes and narrow line equivalent widths
  and line flux-ratios.  We find that the disk-emitters have medium
  luminosities (\about 10$^{44}$\,erg\,s$^{-1}$) and FWHM on average
  six times broader than the AGN in the parent sample.  The
  double-peaked AGN are 1.6 times more likely to be radio-sources and
  are predominantly (76\%) radio quiet, with about 12\% of the objects
  classified as LINERs.  Statistical comparison of the observed
  double-peaked line profiles with those produced by axisymmetric and
  non-axisymmetric accretion disk models allows us to impose
  constraints on accretion disk parameters. The observed H$\alpha$
  line profiles are consistent with accretion disks with inclinations
  smaller than $50^{\circ}$, surface emissivity slopes of 1.0--2.5,
  outer radii larger than \about 2000$R\si{G}$, inner radii between
  200--800$R\si{G}$, and local turbulent broadening of
  780--1800\,km\,s$^{-1}$. The comparison suggests that 60\% of
  accretion disks require some form of asymmetry (\emph{e.g.},
  elliptical disks, warps, spiral shocks or hot spots).

\end{abstract}    

\keywords{\sc{active galactic nuclei, accretion disks, broad line 
emission, emission line profiles}}

\section{Introduction}

Together with a strong continuum emission across the electromagnetic
spectrum from radio to $\gamma$-rays, broad emission lines are one of
the defining characteristics of activity in galaxies. Their prominence
in the ultraviolet (UV) and optical spectra of active galactic nuclei
(AGN), proximity to the central engine, and short timescale
variability, make them good candidates for tracing the gravitational
potential of the central supermassive black hole (SBH) and the
interaction of various distinct kinematic structures in the central
region.  The availability of high signal-to-noise ratio,
high-resolution optical spectroscopy for large samples of AGN
increases their importance as a central region diagnostic.  The
physical structure, dynamics, and luminosity of the region surrounding
the SBH is probably determined solely by the accretion rate of
material in the presence of magnetic fields, the mass of the black
hole and the efficiency with which energy is converted to radiation.
Despite this apparent simplicity, the problem cannot be solved from
first principles, if only because the timescales for radiation
processes are so much shorter than those for hydrodynamical changes.
Our understanding of the influence of magnetic processes and the
interplay of structures on different spatial scales is not adequate to
construct global and self-consistent magneto-hydrodynamic (MHD)
simulations of accretion flows including radiation.

The AGN phenomenon is ultimately dependent on the ability of the gas
to accrete by dissipating angular momentum as it spirals toward the
central source from its reservoir at parsec and sub-parsec scales
\citep[see, for example,][ for a series of lectures on AGN
theory]{Blandford90}.  Either a small amount of material must carry
out the large amount of angular momentum in a powerful wind, allowing
the remaining gas to accrete, or, as we believe is the case, the
material settles in an accretion disk on sub-parsec scales around the
SBH. The disk then mediates the dissipation of angular momentum
through magneto-rotational instabilities \citep{BH91,BH98}. The
presence of disks in accretion powered systems is inevitable
theoretically, and is directly observed in close-contact Galactic
binaries containing a compact object, as well as a in handful of
nearby AGN \citep{4258maser}.

Accretion disk properties are well studied in cataclysmic variables (a
white dwarf in a close binary) and low mass X-ray binaries (a neutron
star or stellar mass black hole in a close binary) through eclipse
mapping, echo mapping and Doppler tomography
\citep{eclipse,lmhb,tomography}.  In these systems detailed
temperature and density distributions of the material in the disk can
be obtained, the disk thickness measured (by its shadow on the
secondary main-sequence companion) and the presence of warps and
spiral structures firmly established \citep{Howell03}.  The orders of
magnitude longer timescales associated with disks around SBHs at the
centers of AGN, their small angular sizes, and the difficulty of
obtaining high signal-to-noise ratio, high resolution, short
wavelength observations, prevent use of these direct techniques to
study the AGN disks. Indirectly, we can use kinematic studies of the
broad emission line region (BLR) to constrain the geometry of material
in the vicinity of the SBH and in rare cases, when the accretion disk
itself clearly contributes to the broad line emission, to investigate
the disk properties using broad, low ionization lines.

Long-term studies of coordinated continuum and line variability
time-delays \citep[``reverberation mapping'',][]{BM82,GHP94} in a few
nearby AGN have shown that high and low-ionization lines follow the
variations in the UV continuum (demonstrating that the line-emitting
gas is photoionized) and that the BLR is optically thick and radially
stratified. Typically the high ionization lines originate closer to
the central black hole than do the Balmer lines \citep[but see the
reverberation studies of 3C390.3,][]{3c390reverb}, and the size of the
BLR is of order light weeks.  Despite significant progress, some
fundamental issues relating to the geometry of the broad line emitting
gas remain unsolved. We do not know whether in general the broad line
region is composed of discrete clouds, winds, disks, or bloated
stellar atmospheres or a combination of these \citep{Korista99}. Broad
line cloud models suffer from formation and confinement problems, and
require too many clouds to reproduce the observed smoothness of the
line profiles \citep{Arav98,Dietrich99}.  The lack of coordinated blue
to red line-wing and line-wing to core variability
\citep{Korista95,Dietrich98,Shapovalova01} suggests that the BLR
material is neither out- nor infalling \citep[but see][for an example
of disk + wind $H\beta$ emission in Mkn 110]{Mkn110}. Disks typically
do not possess enough gravitational energy locally (at distances
inferred from the line widths) to account for the large observed
emission line fluxes \citep{CHF89,Dumont90}.  The fact that all
proposed models suffer drawbacks in certain cases for specific
emission lines supports the idea that the BLR is non-uniform and
suggests the need for fundamentally different components giving rise
to the different broad emission lines, depending on the accretion rate
and central black hole mass. Future studies will thus benefit from
focusing on a well-defined class of AGN and specific broad emission
lines for detailed study.

A small class of AGN, of which almost 20 examples exist in the
literature, shows characteristic double-peaked broad low-ionization
lines attributed to accretion disk emission
\citep{circDisk,EHsample94}. Rotation of the material in the disk
results in one blueshifted and one redshifted peak, while
gravitational redshift produces a net displacement of the center of
the line and distortion of the line profile. A self-consistent
geometric and thermodynamic model can be built, consisting of an
optically thick disk and a central, elevated structure (``ion
torus''), which illuminates the disk, thus solving the local energy
budget problem, while simultaneously accounting for the lack of a
strong big blue bump observed in this class of objects \citep{noBBB}.
Over a decade of line-profile variability monitoring has helped rule
out competitive line emission models for the majority of known
double-peaked AGN \citep[e.g., binary black holes, bipolar outflows,
etc., see][]{BBBreject,Easr98,Easp99} and such work provides a unique
opportunity to study in detail the geometry of accretion together with
the thermodynamic state of the emitting gas in this small sample.

Many open questions concerning the disk emission in this class of rare
AGN remain. Statistical studies are needed to determine how they
differ from the majority of active galaxies, while comparisons of the
observed profiles with disk emission models can constrain the
properties of accretion disks. In this paper we report on the first
statistically large sample (a total of 116 objects included in our
main and auxiliary samples) of H$\alpha$ selected double-peaked AGN
found in spectra taken by the Sloan Digital Sky Survey
\citep[hereafter SDSS,][]{York00}. This is the first in a series of
papers aimed at answering the fundamental questions about the
accretion flow geometry and broad line emission in this class of AGN,
while examining the differences between their overall properties and
those of the majority of active galaxies. Since double-peaked broad
line AGN present some of the most direct evidence for rotation of
material in the vicinity of the supermassive black hole, this large
sample may ultimately provide an explanation for the lack of obvious
disk emission in the majority of AGN.  In this paper we present the
sample selection and line profile measurements in Section 3 after a
brief summary of SDSS observations in Section 2. We comment on the
properties of the double-peaked AGN sample in comparison with the
parent sample of 3126 AGN with $z<0.332$ in Section 4. In Section 5 we
review the accretion disk models and compare the observed H$\alpha$
line profiles to those predicted by axisymmetric and non-axisymmetric
accretion disks in Section 6, followed by summary and discussion.

\section{SDSS Spectroscopic Observations}

The SDSS is an imaging and spectroscopic survey which will image in
drift-scan mode a quarter of the Celestial Sphere at high Galactic
latitudes.  The 2.5\,m SDSS telescope at Apache Point Observatory is
equipped with an imaging camera \citep{Gunn98} with a $3^{\circ}$
field of view which takes 54\,s exposures in five passbands -- \uM,
\gM, \rM, \iM, and \zM~\citep{Fukugita96} with effective wavelengths
of 3551\,\AA, 4686\,\AA, 6166\,\AA, 7480\,\AA, and 8932\,\AA,
respectively.  The identification and basic measurements of the
objects are done automatically by a series of custom pipelines. The
photometric pipeline, {\it Photo} \citep[version 5.3,][]{Lupton03}
performs bias subtraction, flat fielding and background subtraction of
the raw images, corrects for cosmic rays and bad pixels and performs
source detection and deblending. The astrometric positions are
accurate to about $0.1''$ (rms per coordinate) for sources brighter
than $r=20.5$ \citep{SDSSastrom}.  {\it Photo} also measures four
types of magnitudes: point spread function (PSF), fiber, Petrosian,
and model, for all sources in all five bands.  In this paper we use
the model magnitudes, which are \emph{asinh} magnitudes\footnote{The
  inverse hyperbolic sine magnitudes are equivalent to the standard
  astronomical magnitudes for the high signal-to-noise (S/N) ratio
  cases considered here, but are better behaved at low S/N, and are
  well defined for negative fluxes.} \citep{asinhMag} computed using
the best fit surface profile: a convolution of a de Vaucouleurs or
exponential profile with the PSF. The photometric calibration errors
are typically less than $0.03$ magnitudes \citep{SDSSphotom}; see
\citet{photMonitor} for more details on the photometric monitoring.

Using multicolor selection techniques, SDSS targets AGN for
spectroscopy in the redshift range $0< z <5.8$ \citep{Richards02} and
will obtain, upon completion, $10^5$ AGN spectra with resolution of
1800--2100, covering the wavelength region 3800--9200\,\AA. The
apparent magnitude limit for AGN candidates in the lower redshift
($z<2.5$) sample considered here is $i\leq19.1$, with serendipity
objects targeted down to $i\lesssim20.5$.  The serendipity algorithm
\citep{edr} results in a small fraction of AGN targeted irrespective
of their colors, because they match sources in the FIRST
\citep{FIRSTcat} or ROSAT \citep{RASSb,RASSf,Anderson03} surveys.
Additionally, about a third of all $z<0.2$ AGN considered here were
targeted for spectroscopy as part of the main SDSS galaxy sample with
$r\leq17.77$, corresponding to $i\lesssim17.57$ \citep{Strauss02}. Due
to its large areal coverage, accurate photometry (enabling us to
target AGN effectively even close to the stellar color locus) and
large number of spectra, SDSS is uniquely qualified to find large
numbers of rare AGN.  The spectroscopic observations are carried out
using the 2.5\,m SDSS telescope and a pair of double, fiber-fed
spectrographs \citep{Uomoto03}.  Spectroscopic targets are grouped
into $3^\circ$ diameter custom-drilled ``plates'', with 640 optical
fibers each. The fibers subtend $3''$ on the sky, and approximately 80
of them are allocated to AGN candidates on each plate \citep{tiling}.
The $3''$ fiber is equivalent to \about 10\,kpc at $z\sim0.3$, which
will prevent us from selecting low-strength broad emission lines
against a strong stellar continuum \citep{ngc4450,pictorA,Hao03}.
Typical exposure times for spectroscopy are 45--60 minutes and reach a
signal-to-noise ratio of at least 4 per pixel (a pixel $\approx$ 1\AA)
at \gM$= 20.2$.

The spectroscopic pipeline \citep{specPipe} removes instrumental
effects, extracts the spectra, determines the wavelength calibration,
subtracts the sky spectrum, removes the atmospheric absorption bands,
performs the flux calibration, and estimates the error spectrum.  The
spectroscopic resolution is equivalent to $\about 150$\,km\,s$^{-1}$
at H$\alpha$.  The spectroscopic pipeline also classifies the objects
into stars, galaxies, and broad-line AGN while determining their
redshift through $\chi^2$ fits to stellar, galaxy or AGN templates.

For more details concerning the SDSS photometry and spectroscopy and
the various measurement techniques and quantities, we refer the reader
to the Early Data Release publication by \citet{edr} and the Data
Release One publication by \citet{dr1}.

\section{Analysis of the H$\alpha$ Line Region and Sample Selection}

\subsection{Selection of the Double Peaked Sample}

Our H$\alpha$ selection procedure has two steps, the first of which
separates the unusual (predominantly broad and/or asymmetric) from the
symmetric lines\footnote{This step will also select \emph{symmetric}
  double peaked profiles, although in practice these are rare.} using
spectroscopic principal component analysis (PCA), while the second
makes use of multiple Gaussian fitting to distinguish between the
double-peaked and single-peaked asymmetric lines.

\begin{figure}
\plotone{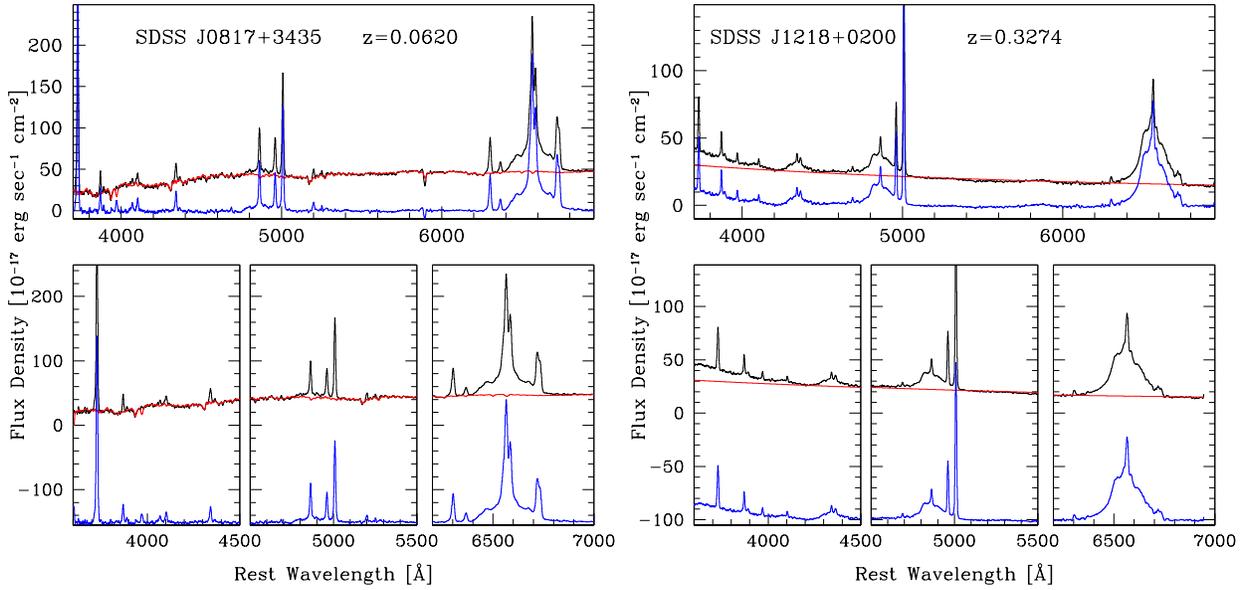}
\caption{Stellar continuum subtraction. The original spectra are in 
  black, the continuum templates are in red, and the residual emission
  lines are given in blue. The spectra shown are the observed flux
  densities, redshifted to $z=0$, vs. the the rest frame wavelength.
  The spectrum on the left is dominated by the galaxy continuum, while
  the one on the right is fit by a simple power law.  The lower panels
  give close-ups of the [\ion{O}{2}]$\lambda$3727, H$\beta$ and
  H$\alpha$ emission line regions; note that here the residual
  emission lines are displaced by $-150$(left) and $-100$(right) units
  for clarity.
\label{galaxy_sub}} 
\end{figure}

The initial sample consists of 5511 objects with $z<0.5$, observed by
SDSS as of June 2002 and classified as AGN by the spectroscopic
pipeline.  A subsample of 3216 AGN (hereafter the ``parent'' sample)
with signal-to-noise ratio per pixel in the red continuum of
$S/N\geq7$, full coverage over the H$\alpha$ line region (rest frame
6000--6900\,\AA, an effective $z<0.332$ cut), and no spectral
defects\footnote{Spectral defects like missing signal over a range of
  wavelengths, problems with sky-subtraction, or artificial features
  produced by badly joined blue and red portions of the spectrum,
  refer to commissioning phase spectroscopy which was also included in
  the analysis; the defect rate during normal operations is
  negligible.}  was selected. Before analyzing the H$\alpha$ line
complex we subtract a sum of power law and galaxy continuum from each
spectrum. The galaxy templates used in the subtraction were created by
PCA of a few hundred high signal-to-noise ratio pure absorption line
galaxy spectra observed by the SDSS \citep{Hao03}. Each galaxy
template covers the range 3814--7014\,\AA\ in the galaxy rest frame
and consists of the eight largest galaxy eigen-spectra (in addition to
the power law continuum and the possibility of including an A star
spectrum to accommodate AGN and galaxies with dominant young stellar
populations), with coefficients fit to best reproduce the continuum.
The PCA method of generating stellar subtraction templates has the
advantage of offering a unique solution with the use of only a handful
of eigen-spectra (i.e.  the orthogonality of the eigen-spectra
guarantees uniqueness).  Example continuum subtractions of selected
double-peaked objects with dominant stellar (left panel) or power law
(right panel) continua are shown in Figure~\ref{galaxy_sub}. The power
law fit is weighted toward the H$\alpha$ and H$\beta$ line regions
($\lambda_{rest}>4400$\AA, excluding 150\,\AA\ around H$\alpha$ and
H$\beta$ and 10\,\AA\ around the strong forbidden lines) whenever a
single power law is insufficient to give a good overall fit, as shown
in the right panel of Figure~\ref{galaxy_sub}. The continuum
subtracted spectra are binned to 210\,km\,s$^{-1}$\,pixel$^{-1}$ to
smooth over any small scale features that might influence the broad
line profile fits.

\begin{figure}
\plotone{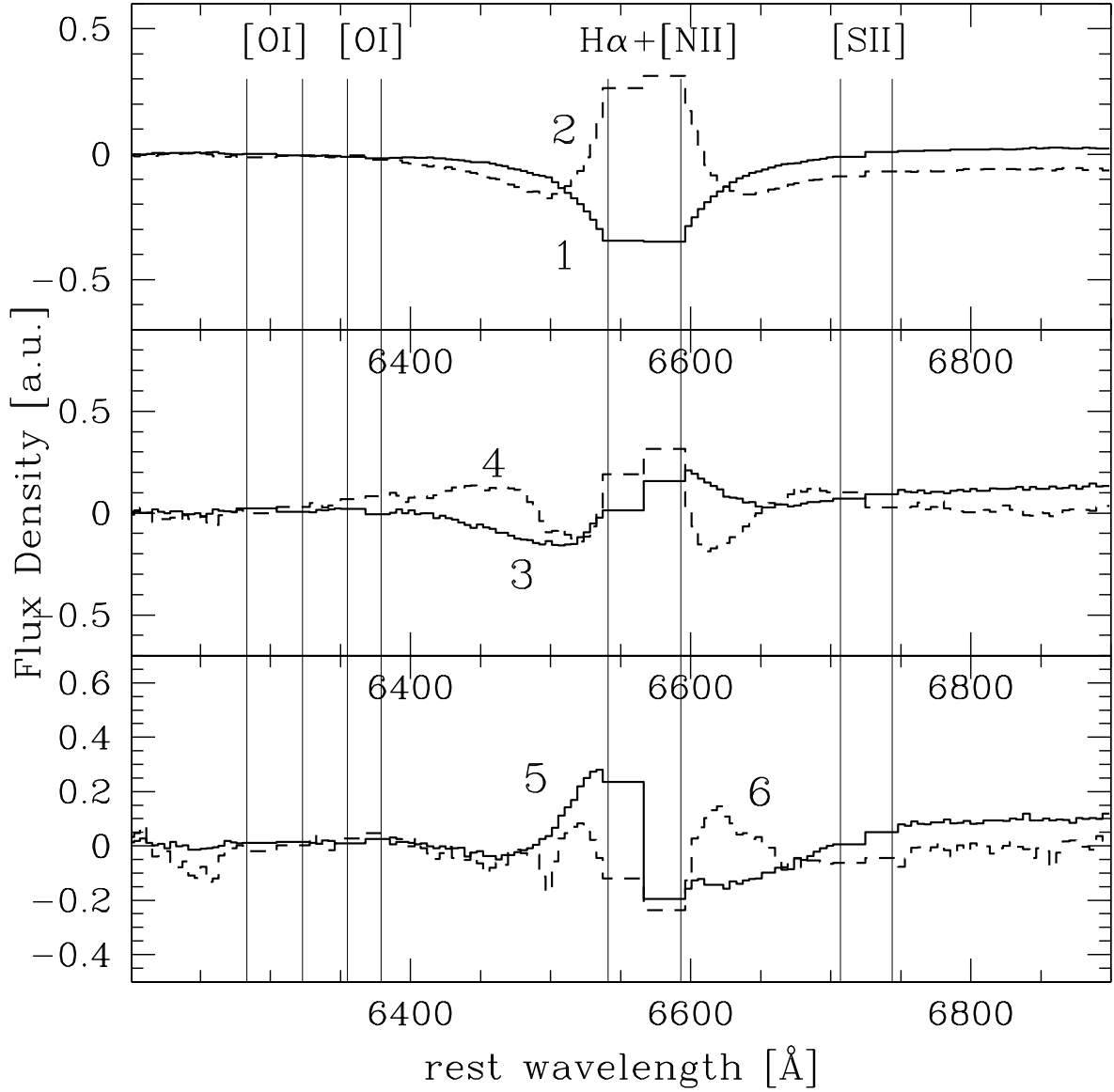}
\caption{The six largest eigen-spectra of the H$\alpha$ line region
  used in the PCA step of the double-peaked AGN selection. The
  vertical lines denote the positions of the narrow line regions which
  were excluded from consideration.
\label{eigen}} 
\end{figure}

\begin{figure}
\plotone{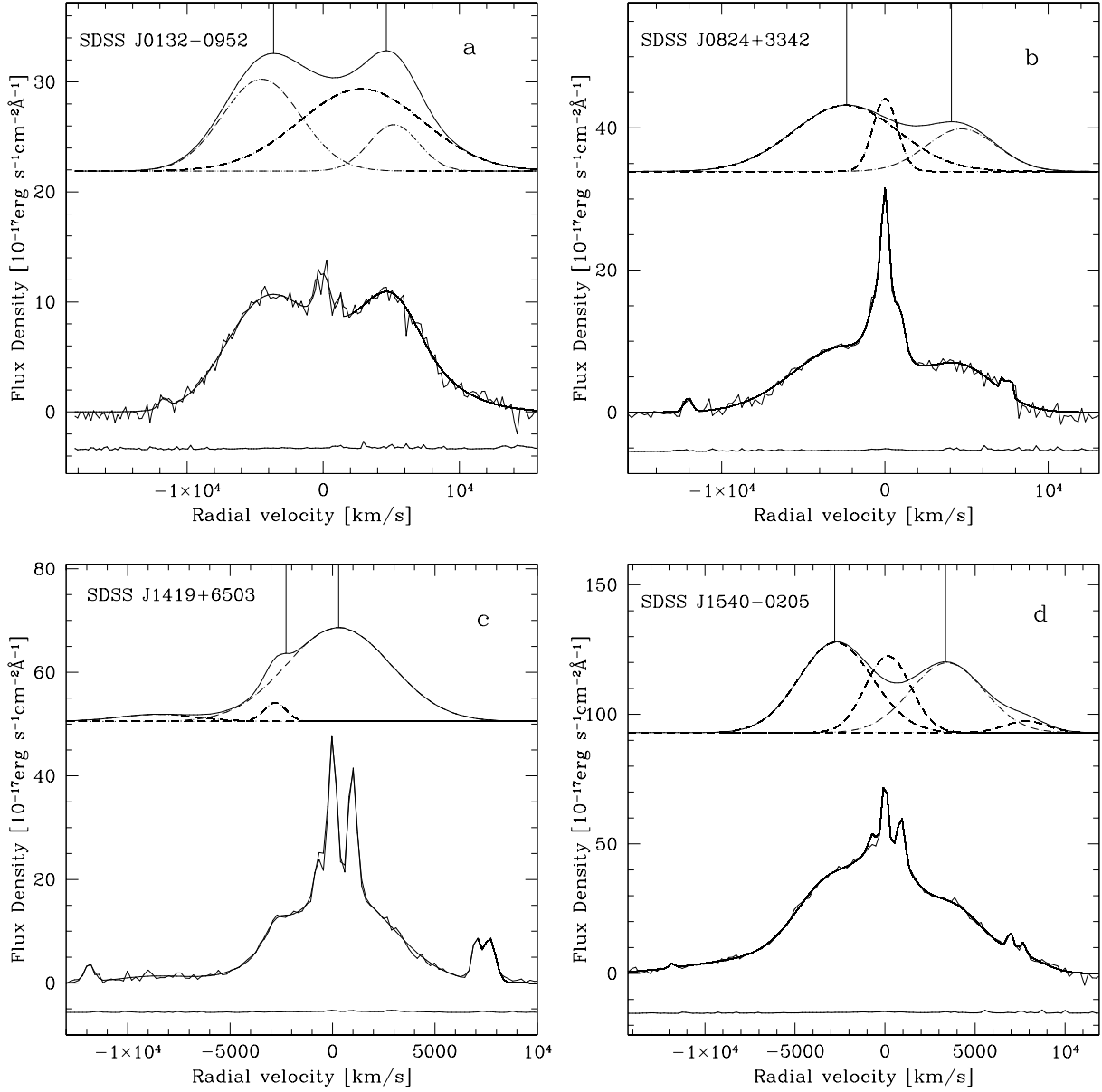}
\caption{Example Gaussian fits. The original spectrum is binned to
  \about 210\,km\,s$^{-1}$ pixels with the sum of Gaussians fit
  overlaid (thick line).  The spectral error is shown below, displaced
  to a negative value (10\% of each plot's maximum flux density) for
  clarity.  The sum of broad components attributed to disk emission
  are shown above with a solid line, while all broad components are
  given separately with dashed lines.  The solid vertical lines from
  the top of each plot point to the positions of the two peaks. a)
  Three broad-component fit. b) Three broad-Gaussian fit, with a
  central broad H$\alpha$ component.  c) Three broad-component fit,
  with the blue peak estimate given by the inflection point of the
  profile at $-2300$\,km\,s$^{-1}$. d) Four broad-component fit with a
  red wing and a broad central H$\alpha$.
\label{gaus_ex_fit}} 
\end{figure}

With 766 of the 3216 AGN visually selected to have symmetric as well
as an unusually large fraction of asymmetric broad line profiles, we
create line-profile eigen-spectra via PCA. The eigen-spectra cover the
6000--6900\,\AA\ rest frame region and use the continuum and stellar
subtracted spectra. The advantage of using eigen-spectra at this step
lies in the fact that a small number of these orthogonal ``principal
components'' can represent the full variance in a sample with no loss
of information. We exclude the narrow line regions from consideration
when creating the eigen-spectra and flux-normalize all lines to
$10^{-16}$ erg\,s$^{-1}$\,cm$^{-2}$, fixing the 6130--6240\,\AA\ 
continuum to zero. Using the eigen-spectra we find the corresponding
coefficients for all 3216 AGN that give good representations of the
original line spectra (the largest 10--15 out of 168 possible
coefficients are usually enough, but we use the first 20 to judge the
goodness of the representation).  Linear combinations of the first,
second, forth and fifth coefficients ($c_1$, $c_2$, $c_4$, and $c_5$,
see Figure~\ref{eigen} for the corresponding eigen-spectra) are then
used to select 645 unusually asymmetric or broad AGN lines out of the
original 3216. The coefficient selection combination, optimal for
selection of an initial visual sample selected by eye by three of the
authors (IVS, NLZ, and MAS) and presented in \citet{Carnegie}, is as
follows:
\begin{equation}
0.64c_1+c_2<-0.084 \quad \wedge \quad 0.55c_2+c_4<0.005 \quad \wedge \quad 1.5c_2+c_5<-0.0045
\label{eqncoeff} 
\end{equation}
At the second selection step, the 645 unusual AGN lines are fit with a
sum of Gaussians: narrow H$\alpha$ $\lambda$6563\,\AA\footnote{Note
  that the wavelengths in SDSS spectra are reported as vacuum
  wavelengths, as detailed in Appendix \ref{sec:vacWave}.},
[\ion{N}{2}]$\lambda\lambda$6548,\,6583\,\AA\ (constrained to 1:3
height ratio and the same width),
[\ion{O}{1}]$\lambda\lambda$6300,\,6364\,\AA, and
[\ion{S}{2}]$\lambda\lambda$6716,\,6731\,\AA, as well as two to four
Gaussians for the asymmetric broad component of H$\alpha$.  Most AGN
line-profiles require more than two broad Gaussians for a good fit to
the broad H$\alpha$ component, as can be seen in the four example fits
of selected double-peaked objects shown in Figure~\ref{gaus_ex_fit}.
The broad component attributed to disk-emission is shown with a solid
line above the spectrum in each case, and consists of three broad
Gaussians in panels a), c), and d) (given separately with dashed
lines) and two broad Gaussians in case b).  A broad central H$\alpha$
component (defined as a component within 5\,\AA\ of the narrow
H$\alpha$ line which is broader than it) is subtracted together with
the narrow lines in panes b) and d), since this emission is likely to
arise in a separate region, as is the case for the prototype
disk-emitter Arp\,102B \citep{uvspec}. Using the sum of broad
Gaussians (less a central broad H$\alpha$ in some cases) we estimate
whether the profile is single or double peaked (see Section
\ref{sec:gaussianfits} for more details).  Because of the large number
of parameters (three per Gaussian), we restarted the
Levenberg-Marquardt fitting procedure \citep{GaussFit} with 10
different initial sets of parameters, taking as our final result the
one with best $\chi^2$. Whenever two of the fits are similar in a
$\chi^2$ sense, but the peak finding method finds two peaks in one
case and only one in the other, the candidate is not considered a
double-peaked AGN, with the exception of 10 interesting cases that are
retained in an auxiliary sample as detailed in Appendix
\ref{sec:Auxiliary}. The sum of broad components of the fit (excluding
any broad central component) is later used in
Section~\ref{sec:gaussianfits} for measuring a number of
line-characterizing quantities for comparison with models.

\begin{figure}
\plotone{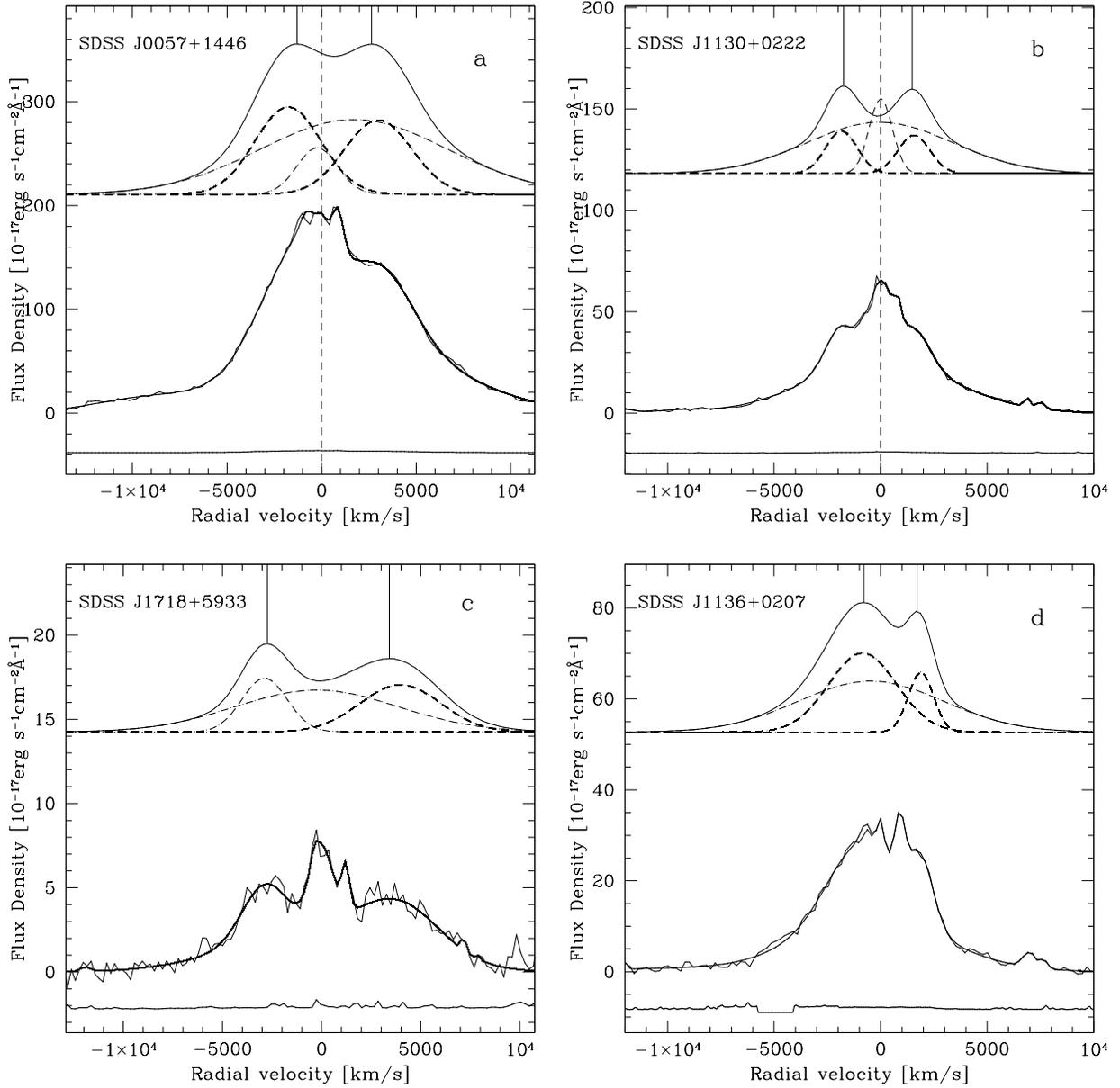}
\caption{Examples of H$\alpha$ selected disk-emission candidates from each
  type: a) strong red shoulder (RS), b) strong blue shoulder (BS), c)
  two separate peaks (2P), and d) blended peaks (2B). The plot details
  are as in Figure~\ref{gaus_ex_fit}.
\label{selectTypes}} 
\end{figure}

Table~\ref{tab1} lists all 85 disk-emitting candidates selected in the
two-step procedure, arranged in order of increasing RA. Additional 31
objects of interest that were not identified by one of the steps of
the algorithm but have interesting line profiles suggestive of disk
emission are presented in Appendix \ref{sec:Auxiliary} and listed
separately in Table~\ref{tab2}\footnote{The original spectra for all
  116 selected disk-emission candidates, the Gaussian fits to the
  H$\alpha$ line region of all 138 exposures (including repeat
  observations) and figures showing all fits are available upon
  request to the first author.}.  The first column in both tables
lists the official SDSS name in the format ``SDSS
Jhhmmss.s$\pm$ddmmss.s'', J2000; we will shorten this to ``SDSS
Jhhmm$\pm$ddmm'' for identification beyond these tables.  The second
column gives the redshift, measured at the [\ion{O}{3}]$\lambda$5007
line as appropriate for the AGN host; in all line profile discussions
below we use velocity coordinates with respect to the narrow H$\alpha$
line.  Columns three through seven in Tables \ref{tab1} and \ref{tab2}
give the apparent model magnitude\footnote{The version of the
  processing pipeline available at the time of writing ({\it Photo}
  version 5.3) systematically underestimates (\emph{i.e.} they are too
  bright) the model magnitudes of galaxies brighter than 20th
  magnitude by 0.2 magnitudes \citep{dr1}; the corrected magnitudes
  will be published as soon as they are available.} of the active
galaxy in the SDSS \emph{ugriz} passband system \citep[corrected for
Galactic extinction, following][]{SFD98} and the last column contains
selection comments.  The objects in Table~\ref{tab2} (hereafter the
``auxiliary'' sample) will be considered as an extra sample treated
separately from the uniformly selected sample of 85 objects (hereafter
``main'' sample) of Table~\ref{tab1} in all subsequent discussions in
this paper.  The 85 objects of Table~\ref{tab1} are classified into
five groups: prominent red shoulder objects (``RS''), prominent blue
shoulder objects (``BS''), two prominent peaks (``2P''), blended peaks
(``2B''), or complex many-broad Gaussian line (``MG''). A ``+C''
indicates that a central broad H$\alpha$ component is included in the
fit.  Figure~\ref{selectTypes} shows example line profiles for each of
the first four types.  In addition a comment ``RL'' marks the
radio-loud AGN (see Section \ref{sec:radio}) in both Tables~\ref{tab1}
and \ref{tab2}.

Twenty-two AGN were observed repeatedly, some multiple times. One of
the reasons for repeat spectroscopic observations was to achieve
acceptable signal-to-noise ratio on early plates, hence not all the
data are of good quality.  Nonetheless, we have 30 repeat observations
of 22 of the candidate disk-emitting AGN with separations ranging from
3 days to 2 years. The repeat exposures are indicated as
``repeat$N$,$P\si{min}$-$P\si{max}$'' in Tables~\ref{tab1} and
\ref{tab2}, where $N$ is the number of repeat observations, excluding
the principal one, and $P\si{min}$ and $P\si{max}$ are the minimum and
maximum separations, in days (for $N=1$ the format reduces
to``repeat1,$P$'').

\begin{figure}
\plotone{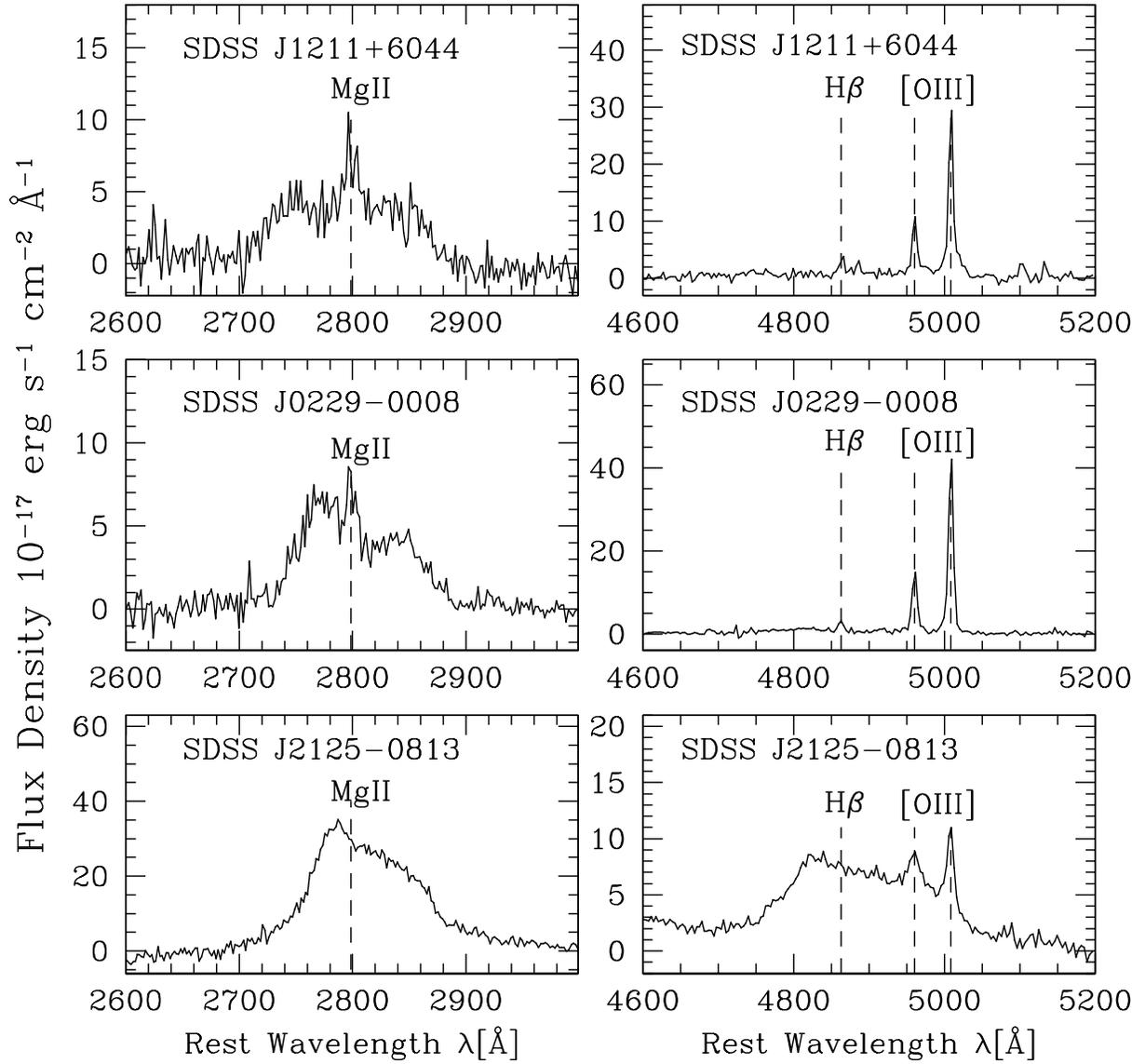}
\caption{Low ionization lines of three disk-emission candidates selected 
  based on their \ion{Mg}{2} and H$\beta$ (bottom right) line
  profiles. The spectra are smoothed to \about 500 km/s. Note the
  broad Fe emission blueward of H$\beta$ in the bottom right panel.
\label{MgIIHbselect}} 
\end{figure}

Disk-emission candidates are selected based on their characteristic
double-peaked H$\alpha$ line profiles. H$\beta$ line profile selection
could, in principle, permit extension of the sample to higher
redshifts, but isolating the broad H$\beta$ component is difficult.
One reason is that the region around H$\beta$ is sometimes heavily
contaminated by Fe emission while in other cases the broad line
component is very weak or completely absent from the spectra
(\emph{i.e.} in objects with a large Balmer decrement), even when
H$\alpha$ is clearly double-peaked.  Selection based on the shape of
\ion{Mg}{2}\,$\lambda$2800 is similarly ambiguous since \ion{Mg}{2}
sits on a pedestal of Fe emission; moreover narrow self-absorption at
the systemic redshift is often present, making it even harder to use
\ion{Mg}{2} for identification of disk-emitters.  Nonetheless we have
three interesting objects which were found by chance based solely on
their \ion{Mg}{2} and H$\beta$ profiles (see
Figure~\ref{MgIIHbselect}).

\subsection{Gaussian Fits and Line Profile Measurements}
\label{sec:gaussianfits}

In order to compare statistically the observed line profiles with
theoretical profiles, we measure a series of profile-characterizing
quantities which are related to one or more of the model disk
parameters as described in Section 5 below. The sum of broad Gaussians
(excluding any central broad H$\alpha$ component) fitted to each
candidate disk-emission AGN provides us with smooth representations of
the observed profiles ideal for such measurements.  Using these smooth
profiles we measure the full width at half maximum (FWHM), the full
width at quarter maximum (FWQM), their respective centroids FWHMc and
FWQMc, the positions of the blue and red peaks ($\lambda\si{red}$ and
$\lambda\si{blue}$), and the blue and red peak-heights ($H\si{red}$
and $H\si{blue}$). All positional measurements (FWHM, FWQM, FWHMc,
FWQMc, $\lambda\si{red}$, and $\lambda\si{blue}$) are quoted in
km\,s$^{-1}$ with respect to the narrow H$\alpha$ line position; the
peak heights ($H\si{red}$ and $H\si{blue}$) are given in $10^{-17}$
erg\,s$^{-1}$\,cm$^{-2}$\,\AA$^{-1}$.  The peak positions are found by
requiring a first derivative numerically close to zero and negative
second derivative.  If this criterion fails and the two peaks are not
obvious in the sum of broad Gaussians (for example, for strongly
one-shouldered objects), we use an inflection point in a few cases to
stand for the second peak (see panel c of Figure~\ref{gaus_ex_fit} for
an example).  By relaxing the numerical threshold for the inflection
point search at this step we can also estimate the peak positions for
most cases of auxiliary sample AGN flagged as ``NotGausSelect'' in
Table~\ref{tab2}. The line widths at half and quarter of the maximum
are measured by first finding the two points on each side of the
profile numerically closest to the desired fraction of the maximum,
and then linearly interpolating between them to find the precise
position.  Table~\ref{tab3} gives the H$\alpha$ measurements for both
the main and auxiliary samples of disk-emission candidates, with
separate measurements at each epoch for objects observed more than
once (a total of 138 entries for the 116 objects).  

\begin{figure}
\epsscale{0.8}
\plotone{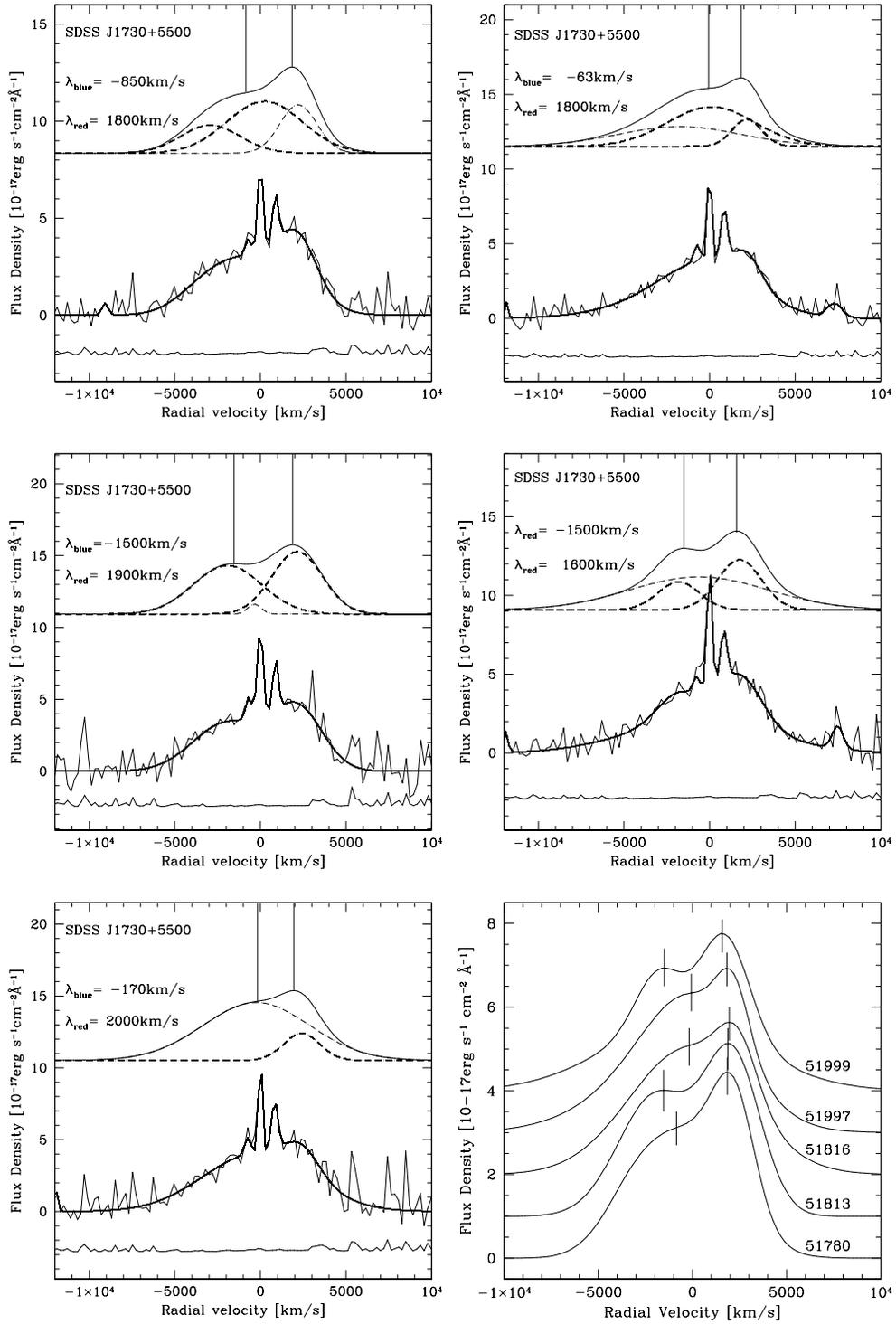}
\caption{Example H$\alpha$ profile variation of a single object over 
  5 epochs, and corresponding line profile measurements. Successive
  observations (from top left down and from top right to the middle
  panel) show small variations in the line profiles on scales of 2-3
  days to 219 days. \emph{Bottom right:} The broad line profiles
  (displaced by 1 unit for clarity and scaled to the same line flux)
  for all 5 observations with the MJD indicated on the right. Note the
  large variations in the measured blue peak position.
\label{profVar}} 
\end{figure}

We measure the line-characterizing quantities for each observed epoch
separately instead of combining the profiles, for two reasons. For
repeat observations over short time intervals ($<$6 months up to a
year), no strong variation in the line shape is expected, and we can
use these profiles separately to quantify the errors of the selection
and line-parameter measurement algorithms.  Whenever the interval
between observations is on the order of years --- the timescale for
which substantial changes in the profile are expected \citep{var} ---
combining the profiles will result in loss of information.
Figure~\ref{profVar} gives example line profile fits to repeat
observations of the same object over half a year. The variations in
the line profile are not large, but are significant (see bottom right
panel of Fig.~\ref{profVar}), and result in substantial difference in
the measured blue peak position. 

Broad line variability in AGN can be caused by changes in the
illuminating continuum \citep{GHP94}, the dynamics of the emitting gas
\citep{var}, or changes in the BLR structure. The timescales of
interest to us are the light-crossing and dynamical timescales, since
they are sufficiently short to produce visible profile change in our
sample.  Changes in the illuminating continuum (reverberation) which
cause changes in the observed line flux, but not the line
shape \citep{WP96}, are apparent on light-crossing timescales
($\tau$\si{light}$=R/c$), which are of order 6--60 days for a
$10^8$--$10^9$M$_{\odot}$ black hole and a BLR size of $R=10^3R_G$
(where M$_{\odot}$ is the mass of the Sun and $R_G=GM/c^2$ is the
gravitational radius around a black hole of mass $M$, $G$ is the
gravitational constant and $c$ the speed of light). If the accretion
disk is non-axisymmetric, the disk inhomogeneities will orbit on
dynamical timescales --- $\tau$\si{dynamical} \about\ 6\,$M\,R^{3/2}$
months --- causing changes in the line shape.

\subsection{Error Estimates}
\label{sec:Merrors}

\begin{figure}
\epsscale{0.6}
\plotone{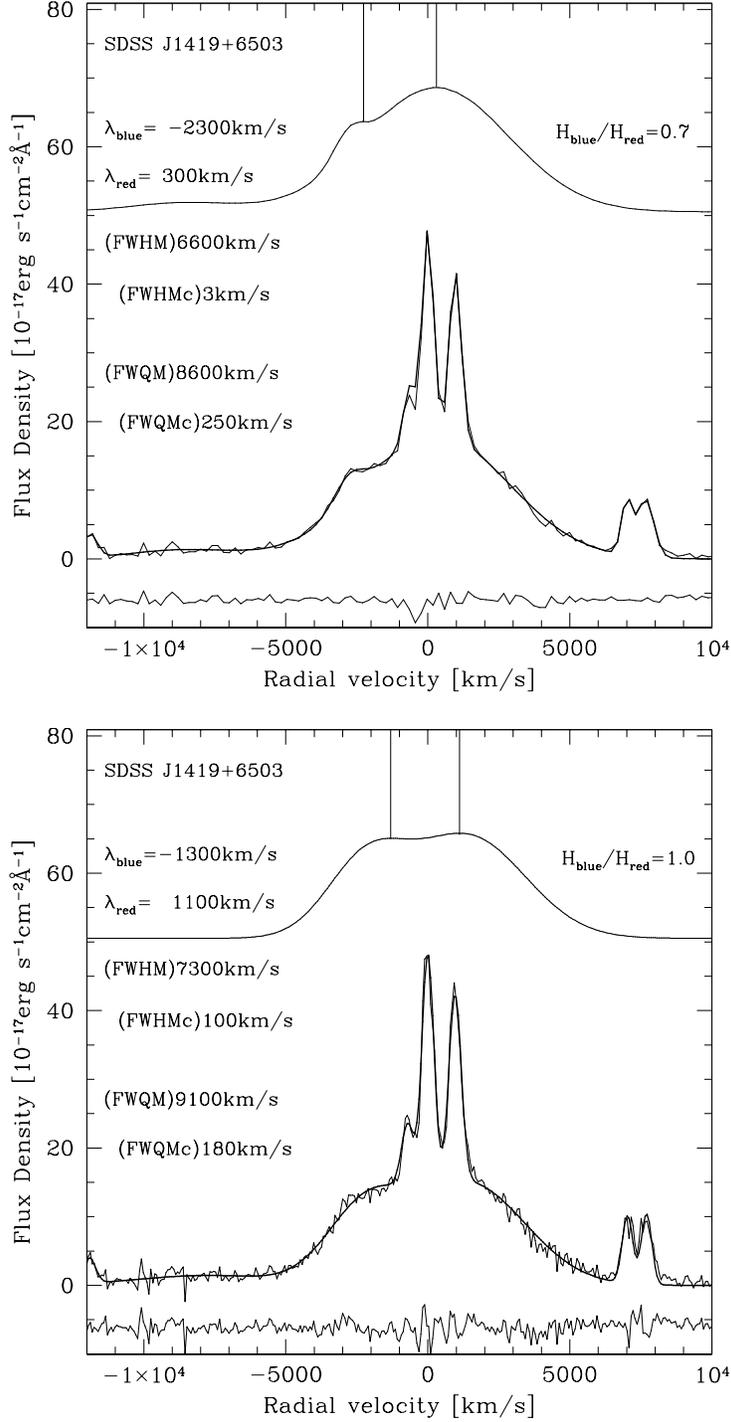}
\caption{Line-parameter variation due to alternative processing.
  The profile measurements of a line with an alternative fit (bottom
  panel) results in notably different red and blue peak positions and
  heights from these in the final analysis (top panel). The plot
  details are as in Figure~\ref{gaus_ex_fit}, except the residuals
  replace the error spectra here and the line-profile measurements are
  listed in each case. The top panel fit is slightly better in a
  $\chi^2$ sense and is preferred.
\label{measureVar}} 
\end{figure}

\begin{figure}
\epsscale{1}
\plottwo{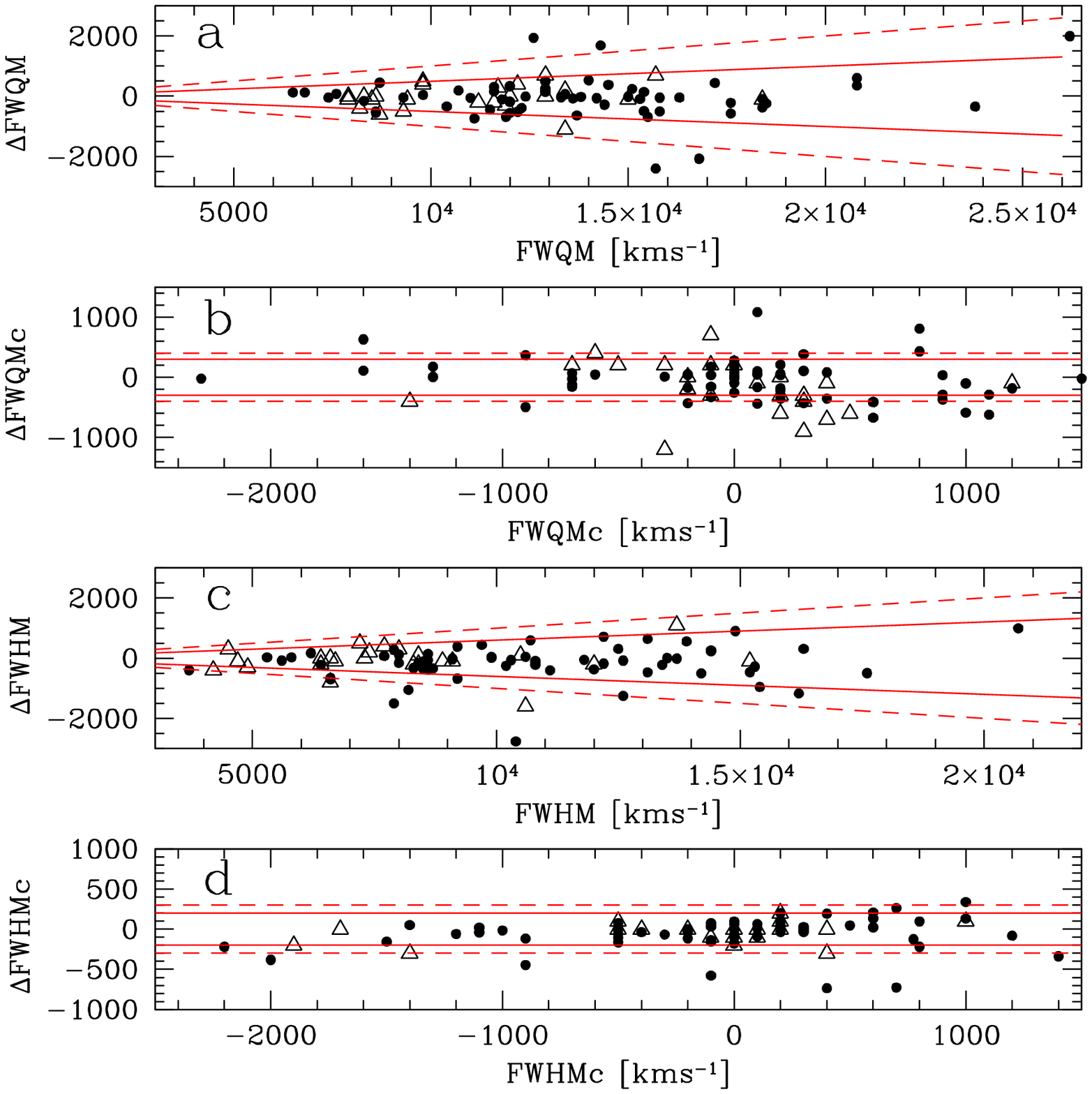}{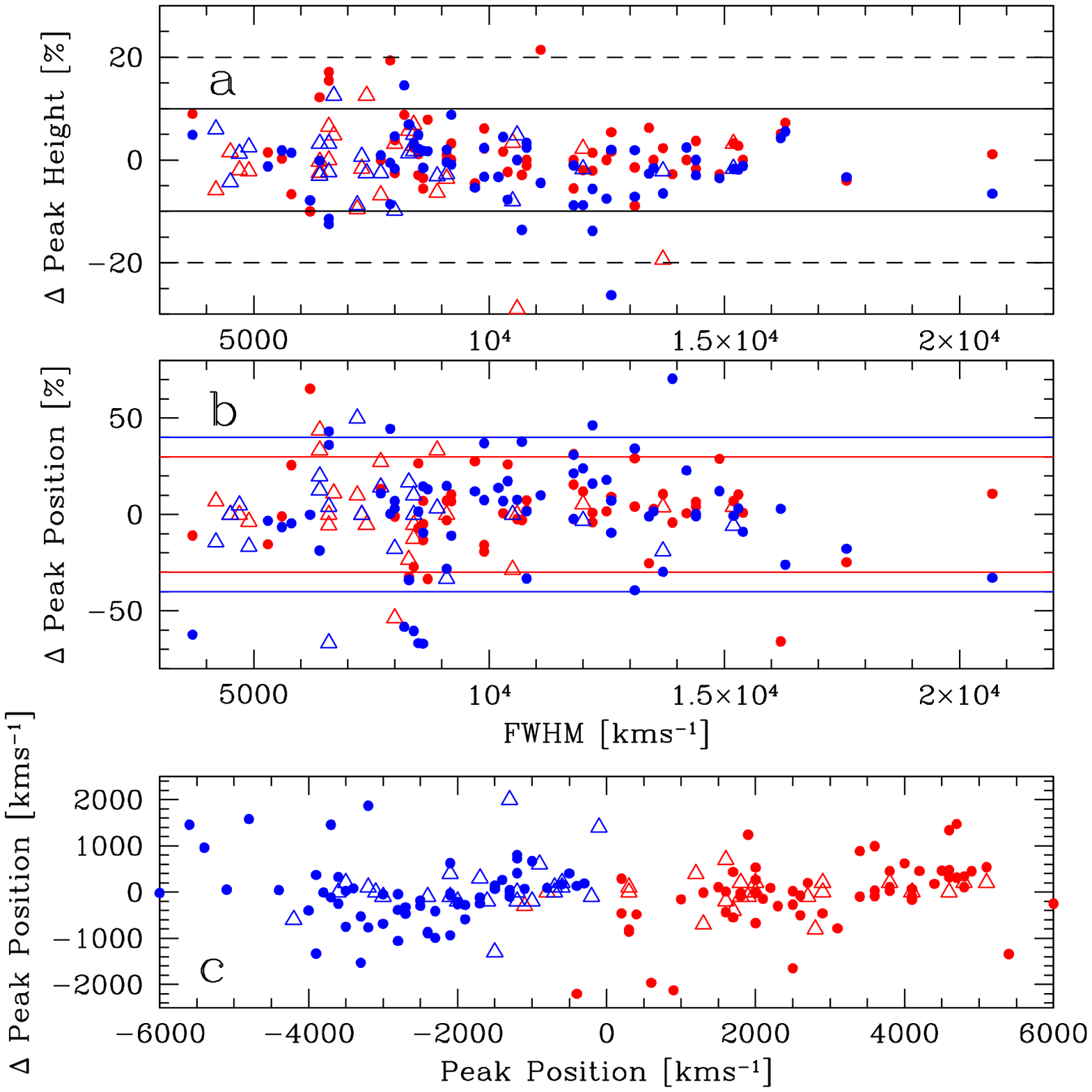}
\caption{Error estimates for the measured line parameters
  from the reprocessed subsample (filled circles) and the repeat
  observations of the same AGN (triangles). 80\% of the measured
  parameter variation of the combined subsamples is enclosed within the
  solid lines on each plot, 90\% within the dashed lines. {\it Top:}
  a) $\Delta$ FWQM (in kms$^{-1}$) vs. FWQM, the solid and dashed
  lines correspond to 5\% and 10\% error, respectively. b) $\Delta$
  FWQM centroid (in kms$^{-1}$) vs. FWQM centroid. c) $\Delta$ FWHM
  (in kms$^{-1}$) vs. FWHM, the solid and dashed lines correspond to
  6\% and 10\% errors. d) $\Delta$ FWHM centroid (in kms$^{-1}$) vs.
  FWHM centroid. {\it Bottom:} a) Percent difference in measured
  heights for the red (red symbols) and blue (blue symbols) peaks vs.
  FWHM.  b) Percent difference in measured peak positions vs. FWHM. c)
  Difference in peak positions vs. peak position.
\label{errPlot}} 
\end{figure}

Errors in the final line parameter measurements can occur in any of
the processing steps --- the continuum subtraction, the Gaussian
fitting, or the line parameter estimation. Although the line
measurements of the Gaussian fits are very precise, some complex line
profiles allow alternative peak representations --- for example in
cases with only one prominent shoulder (see Fig.~\ref{measureVar} for
an extreme example).  In order to estimate the fitting errors we
compare the line parameter measurements obtained in an earlier
separately processed subsample of 63 line profiles (hereafter
``reprocessed'' sample) with those of our final sample given in
Table~\ref{tab3}. We also compare the differences in line measurements
between repeat observations over timescales of less than a year (24
cases), after normalizing the lines to a common total flux.  The two
methods give similar results, although they test for discrepancies
coming from algorithm issues in the first case, and S/N ratio and
repeatability in the second.  The earlier processed subsample uses the
unbinned original spectra for the Gaussian fits, has a slightly older
version of the continuum subtraction algorithm and performs only one
Gaussian fit, instead of restarting the fits with different initial
values.  The differences in line measurements found by both methods in
each of the 2 subsamples (the reprocessed subsample and the repeatedly
observed one) are given in Figure~\ref{errPlot}. We define the errors
as the limits containing 80\% of the combined subsamples' variation;
for example, the FWQM measurement differences are within 5\% of the
value measured in km\,s$^{-1}$ for 80\% of the combined subsamples.
The FWHM measurements (in km\,s$^{-1}$) have \about 6\% error, the
FWHM centroids about $\pm200$\,km\,s$^{-1}$, the FWQM centroids about
$\pm300$\,km\,s$^{-1}$, the peak height errors are \about 10\% of the
flux density measured in units of $10^{-17}$
erg\,s$^{-1}$\,cm$^{-2}$\,\AA$^{-1}$, the peak positions between 30\%
(red peaks) and 40\% (blue peaks) of the measured position in
km\,s$^{-1}$.  The errors in all measured quantities are recorded in
Table \ref{tab4}.

\section{Double-peaked AGN Sample Properties}

\subsection{Emission Line Widths and Strengths}

\citet{EHsample94}, \citet{Easr98}, and \citet{ngc4450} define a set
of characteristics of the double-peaked disk-emitters which
distinguish them from the majority of AGN. In their studies of
predominantly lobe dominated, radio-loud AGN they find that, \emph{on
  average}, the disk emitters have broad lines that are twice as broad
as the typical radio galaxy or radio-loud AGN, have large
contributions of starlight to their continuum emission, high
equivalent widths of low-ionization forbidden lines like
[\ion{O}{1}]$\lambda$6300, and [\ion{S}{2}]$\lambda\lambda$6716,\,6731
(about twice that of average AGN), and line ratios,
[\ion{O}{1}]$\lambda$6300/[\ion{O}{3}]$\lambda$5007 and
[\ion{O}{2}]$\lambda$3727/[\ion{O}{3}]$\lambda$5007, that are
systematically higher. With the exception of the stellar contribution
to the continuum, we investigate these properties for the main and
auxiliary samples in relation to those of the parent sample of 3216
AGN with $z<0.332$. Due to the aperture bias resulting from the $3''$
SDSS fibers, the relative stellar contribution we measure would be
biased toward higher values for higher redshift objects, and
consequently we postpone the estimate of stellar contribution to the
continuum until small aperture data is available.

\begin{figure}
\epsscale{1.0}
\plotone{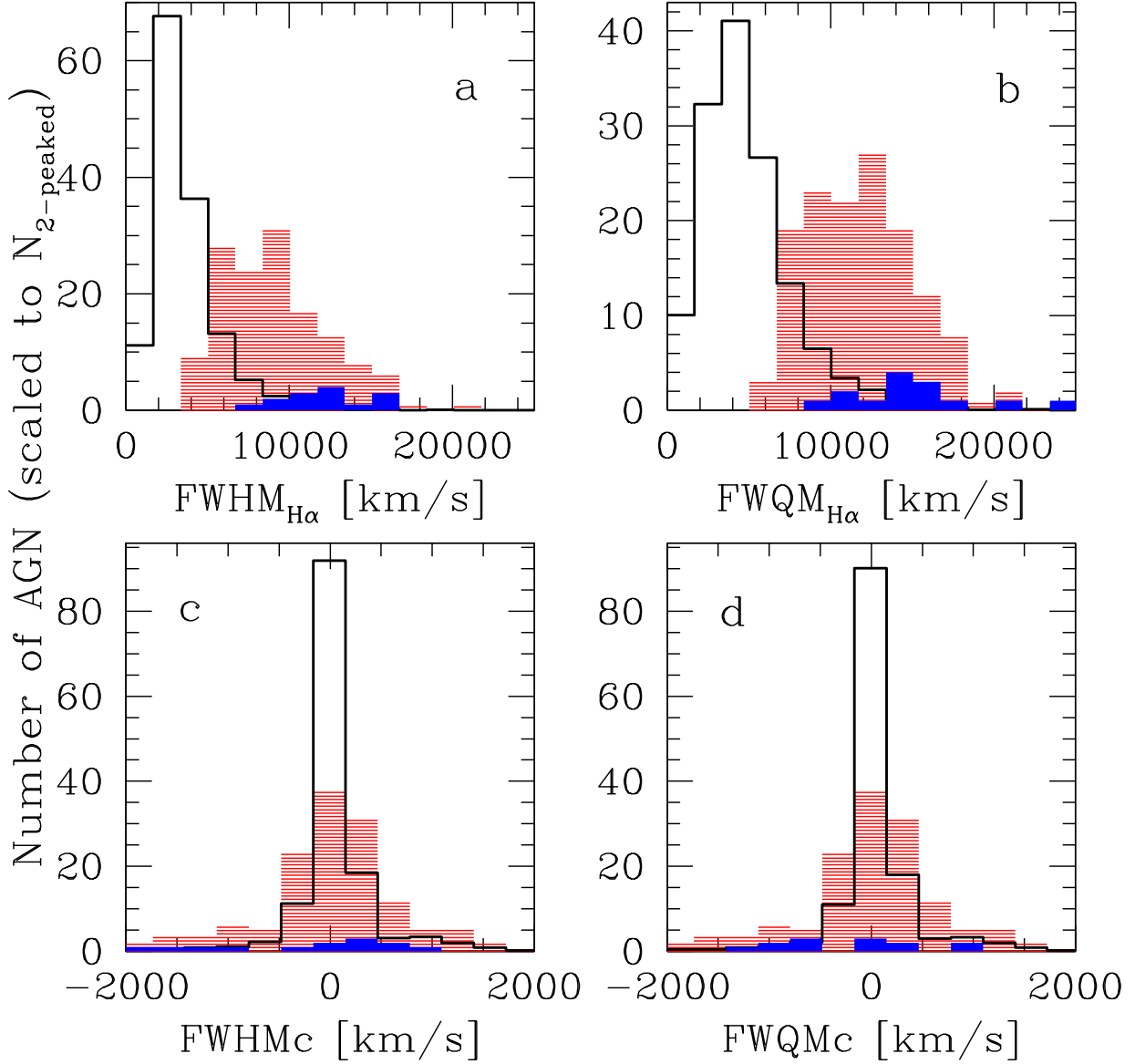}
\caption{Comparison between the distributions of the FWHM (panel a), 
  FWQM (b), FWHM centroid (c), and FWQM centroid (d) of the parent
  $z<0.332$ AGN sample (black hollow histograms) and the selected
  double-peaked emitters from Table \ref{tab3}. The solid blue
  histogram shows the distribution of radio loud double-peaked AGN.
  The total number of AGN have been scaled to the number of
  double-peaked AGN, except for the subsample of radio-loud AGN.
\label{width_centroid}} 
\end{figure}

Figure~\ref{width_centroid} shows the comparison between the selected
double-peaked (red shaded histogram) and parent (black hollow
histogram) AGN samples' FWHM, FWQM and their respective centroids.
The PCA step of the double-peaked AGN selection procedure
preferentially picks up broader than usual H$\alpha$ lines, selecting
about 56\% of all objects with FWQM$>$6000\,km\,s$^{-1}$ (those broad
lines are in turn about one-third of all 3216 AGN lines considered).
The final Gaussian-fit selected disk-emission candidates (shown with
red histogram in Figure~\ref{width_centroid}) have even larger widths,
with 90\% of the main and auxiliary samples having
FWQM$>$8000\,km\,s$^{-1}$; there are only five objects with
FWHM$<$5000\,km\,s$^{-1}$. Thus the PCA selection step is not
restrictive, and we find than the disk-emission candidates have
broader line profiles than the majority of AGN, in agreement with
\citet {EHsample94}. The radio loud subsample, which is better matched
in properties to the sample of known disk-emission AGN, (14 out of our
17 radio loud AGN have Gaussian fit measurements for H$\alpha$, see
Appendix \ref{sec:Auxiliary}) has larger FWHM and FWQM than both the
double-peaked and the parent samples of AGN.

The two lower panels in Figure~\ref{width_centroid} give the FWHM and
FWQM line centroids for the double-peaked (red shaded histogram) and
the parent sample of AGN (black hollow histograms). The two
distributions are significantly different, with the double-peaked AGN
having larger red (and blue!) shifts than the general AGN population.
The larger redshifts of the double-peaked AGN are expected, as they
arise naturally from gravitational redshift of the disk emission in
the vicinity of the SBH. The large blueshifts cannot be explained by
a simple axisymmetric disk emission, but could arise in a
non-axisymmetric disk, as we argue in Section \ref{sec:modComp}. The
solid blue histogram shows 14 out of the 17 radio loud double-peaked
AGN with H$\alpha$ centroid measurements. There is no significant
difference in the distribution of the centroids for the radio loud
double-peaked AGN in comparison with the full AGN sample.

\begin{figure}
\epsscale{1.0}
\plotone{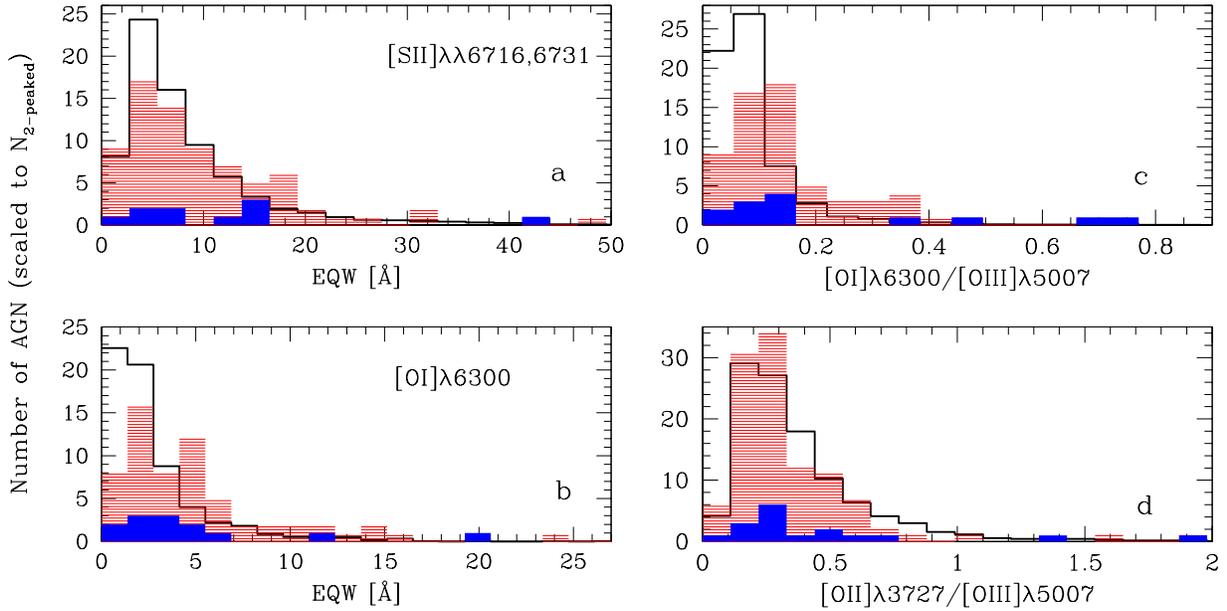}
\caption{Comparison of the equivalent widths (panels a and b) 
  and low-ionization line ratios (c and d) between the parent sample
  of AGN with $z<0.332$ (black histograms) and the selected
  double-peaked AGN (red shading). Note that the two samples are well
  matched in luminosity and redshift. The solid blue histogram shows
  the radio loud subsample from the sample of selected double-peaked
  AGN. The total number of AGN have been scaled to the number of
  double-peaked AGN, except for the subsample of radio-loud AGN.  The
  double-peaked AGN tend to have large equivalent widths of
  [\ion{O}{1}] and higher [\ion{O}{1}]/[\ion{O}{3}] flux ratios.
\label{eqw}} 
\end{figure}

Because the stellar contribution to the continuum changes with
redshift and we measure the line equivalent widths with respect to the
total continuum \citep[stellar + AGN power law, as in][]{EHsample94},
the equivalent width measurements could depend on redshift. We measure
the [\ion{O}{1}]$\lambda$6300 and
[\ion{S}{2}]$\lambda\lambda$6716,\,6731 equivalent widths with respect
to the blue and red continuum in the immediate vicinity (\about 10\AA\ 
on each side) of the lines. No trend of equivalent width with redshift
is apparent.  The distributions of the equivalent widths for the
parent sample of 3216 AGN and the double-peaked AGN (of both main and
auxiliary samples) are presented in Figure~\ref{eqw}.

Overall, a large fraction of the double-peaked sample has
low-ionization line equivalent widths and ratios similar to those of
the parent sample.  The double-peaked AGN tend to have larger
equivalent widths of [\ion{O}{1}]$\lambda$6300 and higher
[\ion{O}{1}]$\lambda$6300/[\ion{O}{3}]$\lambda$5007 flux ratios than
do the majority of AGN. A Kolmogorov-Smirnov (KS) test confirms that
there is a significant difference between the distributions of the
[\ion{O}{1}]$\lambda$6300 equivalent widths and the
[\ion{O}{1}]$\lambda$6300/[\ion{O}{3}]$\lambda$5007 flux ratios for
the double-peaked and parent AGN samples (the null hypothesis, that the
two distributions are the same is rejected with p-values of less than
0.1\%).  There is no strong evidence from Figure~\ref{eqw} or a KS test
that the [\ion{S}{2}]$\lambda\lambda$6716,\,6731 equivalent widths and
[\ion{O}{2}]$\lambda$3727/[\ion{O}{3}]$\lambda$5007 flux ratios of the
double-peaked AGN are different from those of the majority of AGN. The
sample of double-peaked AGN of \citet{EHsample94} is quite small (11
objects fitted with circular disk model) and the spread of measured
equivalent widths is larger than that for their parent sample of 84
radio-loud AGN, which shifts the average of their disk-emission sample
to larger equivalent widths.  The radio loud part of our double-peaked
sample is shaded blue in the histograms of Figure~\ref{eqw}. There is
no significant difference between the equivalent widths and line
ratios of the radio-loud double-peaked AGN in comparison with the full
double-peaked sample (nor is there a trend with the radio-loudness
indicator of any of these quantities), but the number of radio loud
objects (ranging from 10 objects with non-zero
[\ion{S}{2}]$\lambda\lambda$6716,\,6731 equivalent width measurements
to 17 objects with [\ion{O}{2}]$\lambda$3727/[\ion{O}{3}]$\lambda$5007
flux ratio measurements) is too small to state this with confidence.

\begin{figure}
\epsscale{1.3}
\plottwo{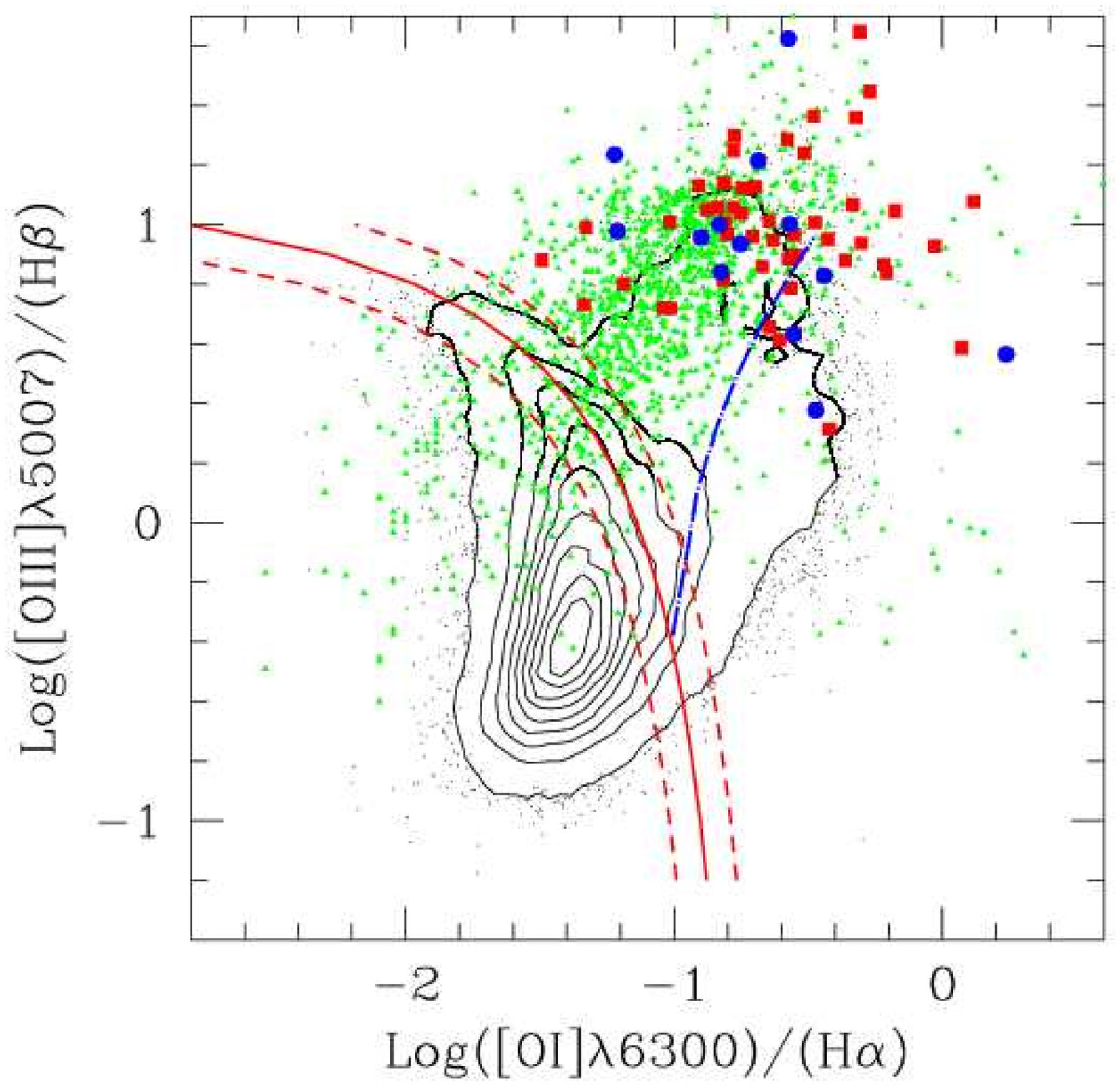}{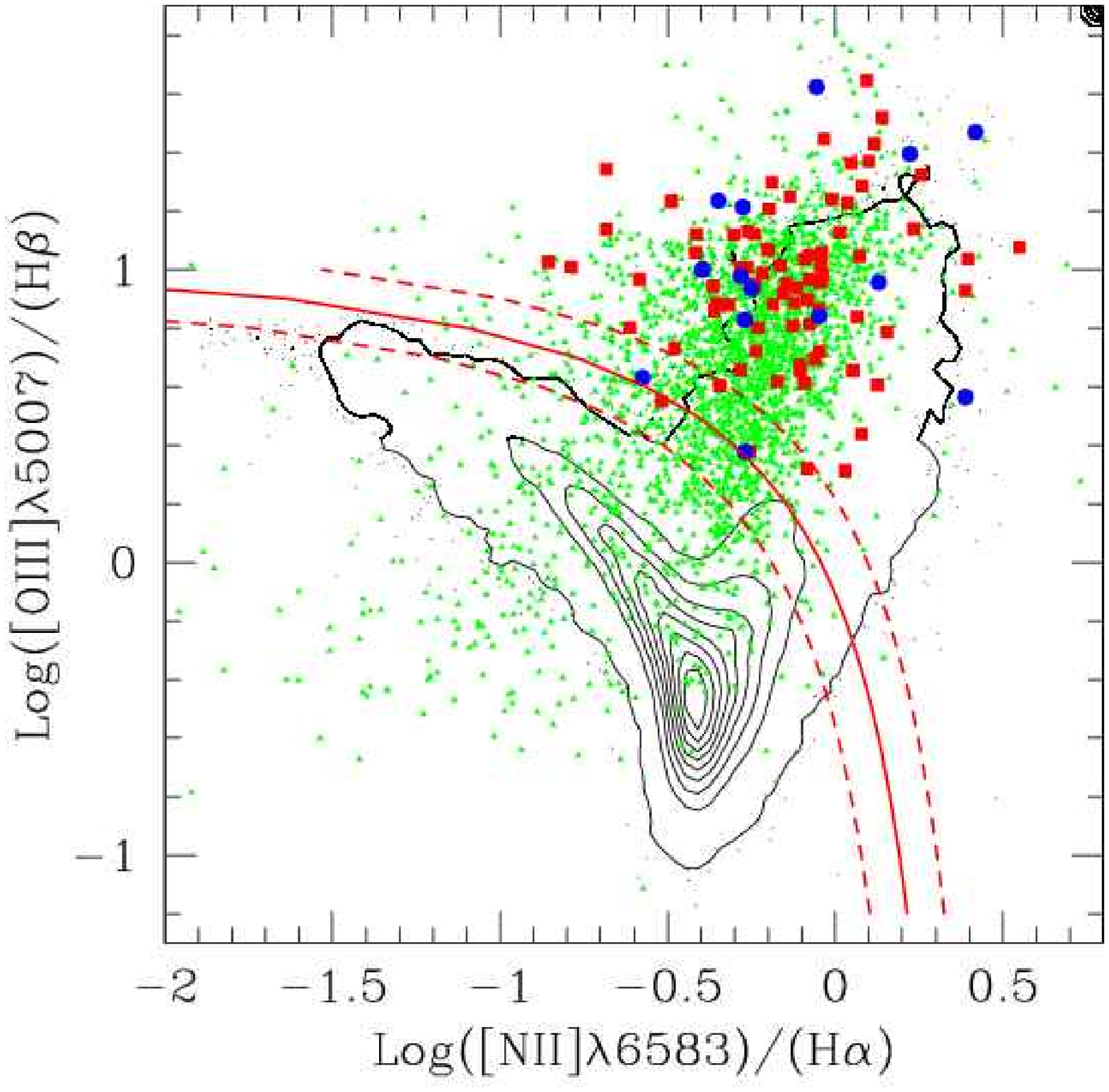}
\caption{Position of double-peaked AGN (red squares) and the parent
  $z<0.332$ AGN sample (green triangles) on narrow line
  Veilleux-Osterbrock diagnostic diagrams. The radio loud AGN from the
  double-peaked sample are shown as solid blue circles. The black
  contours represent over 50,000 Sy2 and star-forming galaxies from
  the \rM$<17.77$ SDSS main sample \citep{Hao03}. The solid red curve
  (the dashed curves correspond to the model uncertainty) is a
  theoretical prediction by \citet{Kewley01}: Sy2s occupy the portion
  of each graph above the red curve, normal star-forming galaxies
  below. The blue dot-dashed curve in the [\ion{O}{1}]/H$\alpha$ vs.
  [\ion{O}{3}]/H$\beta$ plot is Kewley's theoretical prediction
  separating the LINERs (to the right of the line) from the Sy2s.
\label{narrowL}} 
\end{figure}

Figure~\ref{narrowL} shows two of the \citet{VO87} AGN diagnostic
diagrams. The theoretical results of \citet{Kewley01}, given as red
solid lines, separate the Seyfert 2 active galaxies (Sy2s, lying above
the solid red curves in both panels) from the majority of star-forming
galaxies.  The contours represent \about 50,000 galaxies from the main
SDSS sample \citep[with an apparent magnitude cut of
\rM$<$17.77,][]{Strauss02} from \citet{Hao03}, showing the good
agreement between the theoretically computed separation and the data.
The parent sample of more luminous, higher redshift AGN used for this
paper (green triangles) and the sample of double-peaked AGN (red
squares) have similar narrow [\ion{O}{1}]/H$\alpha$,
[\ion{N}{2}]/H$\alpha$, and [\ion{O}{3}]/H$\beta$ flux ratios, with
the [\ion{O}{1}]/H$\alpha$ and [\ion{N}{2}]/H$\alpha$ line-flux ratios
0.2 to 0.3\,dex smaller than the lower luminosity Sy2s.  The errors in
our higher redshift sample narrow-line fluxes introduced by the
subtraction of the broad line component can reach 20-40\%.  There is
no difference between the narrow line ratios of the radio quiet (blue
squares in Fig.~\ref{narrowL}) and radio loud (solid blue circles in
Fig.~\ref{narrowL}) double-peaked AGN.

The dot-dashed blue curve in the [\ion{O}{1}]/H$\alpha$ vs.
[\ion{O}{3}]/H$\beta$ plot in Fig.~\ref{narrowL} is the
\citet{Kewley01} theoretical prediction separating the Sy2s, situated
above and to the left of the curve, from the low-ionization nuclear
emission region (LINER) galaxies to the right and below the curve.
LINERs are believed to be low-luminosity AGN with normal narrow line
regions which either have abnormally low ionization parameters or are
powered by a combination of starburst and AGN. There is some evidence
that LINERs are associated with disk emission, since double-peaked
Balmer line profiles have appeared in well known LINERs NGC 1097,
Pictor A and M81 \citep{ngc1097,Sulentic95,pictorA,M81}. Using
Kewley's criterion (situated below and to the right of the blue curve
in Fig.\,\ref{narrowL}) we find that 12 of the double-peaked AGN at
low redshift are LINERs (\about 12\% of the main and auxiliary
samples).  Uncertainties in the line ratios of 20-40\% (0.1-0.2\,dex)
or in the theoretical curve \citep[\about 0.1\,dex, ][]{Kewley01}
could result in an under- or over-estimate of the fraction of LINERs
(\about 9\% of all selected AGN are within 0.1-0.2\,dex of the
separation line).

\subsection{Colors and Host Galaxies}

\begin{figure}
\epsscale{0.8}
\plotone{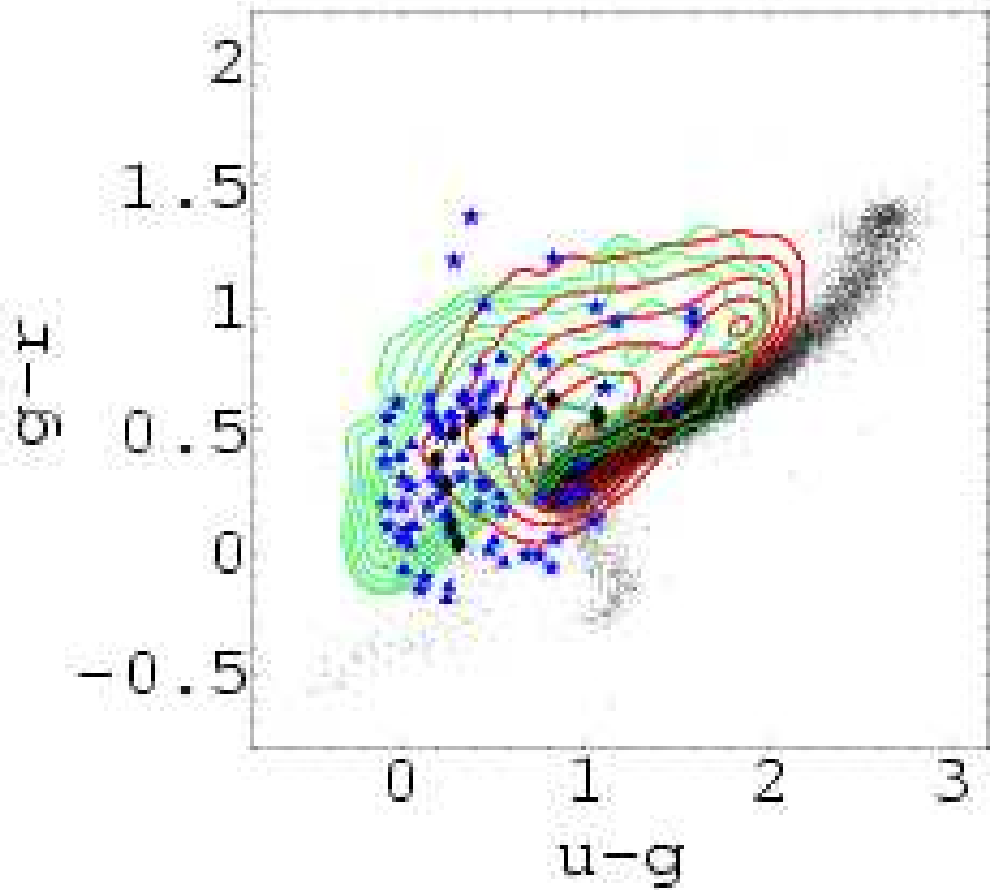}
\caption{AGN host colors (blue stars for the main sample, blue 
  triangles for the auxiliary AGN) compared to stellar colors (black
  dots), galaxy colors (red contours) with the characteristic
  morphological separation into early and late types, and the colors
  of the full AGN sample with $z<0.33$ (green contours). The black
  diamonds (upper right to lower left) are the average colors of AGN
 with redshifts between 0.025 and 0.475 in 0.05 redshift intervals.
\label{agncolors}} 
\end{figure}

The SDSS star-galaxy separation criterion requires that the PSF
magnitude be fainter than the model magnitude by $>0.145$ for extended
objects. The criterion is found to be $>$95\% accurate for objects as
bright as our sample \citep{dr1}. According to this criterion, 32\% of
our full sample (both main and the auxiliary sample) are spatially
unresolved.  About 44\% of the candidates are sufficiently extended to
attempt a crude visual classification; 60\% of these appear to have
early type morphologies (E, SO, Sa).  Figure~\ref{agncolors} gives the
total galaxy+AGN colors for the main and auxiliary samples in
comparison to the colors of stars, galaxies and all AGN with $z <
0.33$. Since the colors of low redshift AGN are quite varied depending
on the host galaxy and the strength of the central nuclear source, the
\uM-\gM\ colors of low redshift sources could be significantly
different from the average colors of either AGN or non-active
galaxies.

\begin{figure}
\epsscale{1.0}
\plotone{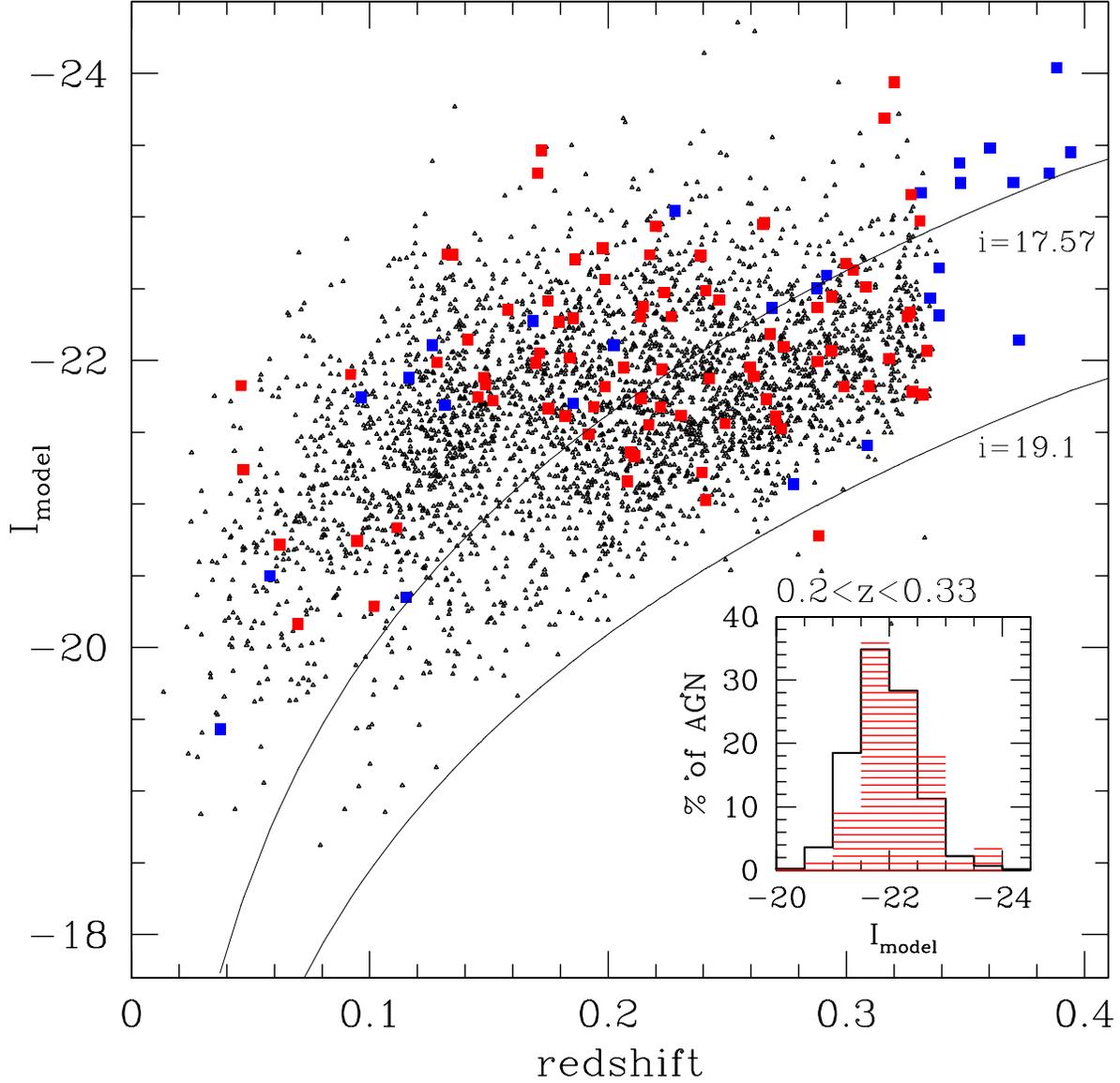}
\caption{Absolute $i$-band magnitude distribution of double-peaked
  AGN from the main (red squares) and the auxiliary samples (blue
  squares) in comparison with that of the original sample of all
  $z<0.332$ AGN (small black symbols). The two curves denote the
  apparent magnitude limits for the AGN (lower curve, $i=19.1$) and
  the galaxy (upper curve, $i\approx$17.57) SDSS target selection for
  spectroscopy.  The inset shows the absolute magnitude histogram for
  the $0.2<z<0.33$ subsamples of the main and original samples.
\label{absmag}} 
\end{figure}

Table~\ref{tab5} lists the total (AGN+host galaxy) luminosities of our
sources in erg\,s$^{-1}$ in the five SDSS bands, computed using
$\Omega_{\lambda}=0.73$, $\Omega\si{m}=0.27$, flat cosmology, with
H\si{o}=72\,km\,s$^{-1}$\,Mpc$^{-1}$. The majority of the candidate
disk-emitters have luminosities of a few $\times 10^{44}$
erg\,s$^{-1}$, similar to the average luminosity of all AGN observed
with the SDSS in the $z<0.332$ redshift range. Figure \ref{absmag}
presents the redshift vs.\ absolute magnitude diagram. The majority of
double-peaked candidates (shown in red for the main sample) have
absolute magnitudes within a magnitude of I$\si{model}=-22$, similar
to the rest of the AGN with $z>0.2$. This is equal to the faintest
magnitude of the standard SDSS definition of a quasars, even though
the model magnitudes used here overestimate the nuclear luminosity by
up to a magnitude for nearby AGN by including the host galaxy
contribution. From Figure~\ref{absmag} it appears that lower redshift
($z\lesssim0.2$) double-peaked AGN tend to be more luminous than the
average AGN, but the numbers are too small to make a statistically
rigorous statement; moreover there are selection effects that affect
the result, since about 1/3 of the AGN at $z<0.2$ were targeted for
spectroscopy as ``main galaxies'' with a brighter apparent magnitude
cut of $r<17.77$ (which corresponds to $i\lesssim17.57$) while for
$z>0.2$ almost all AGN are targeted by the AGN algorithm with a
fainter cut at $i<19.1$.

\subsection{FIRST counterparts}
\label{sec:radio}

\begin{figure}
\epsscale{1.0}
\plotone{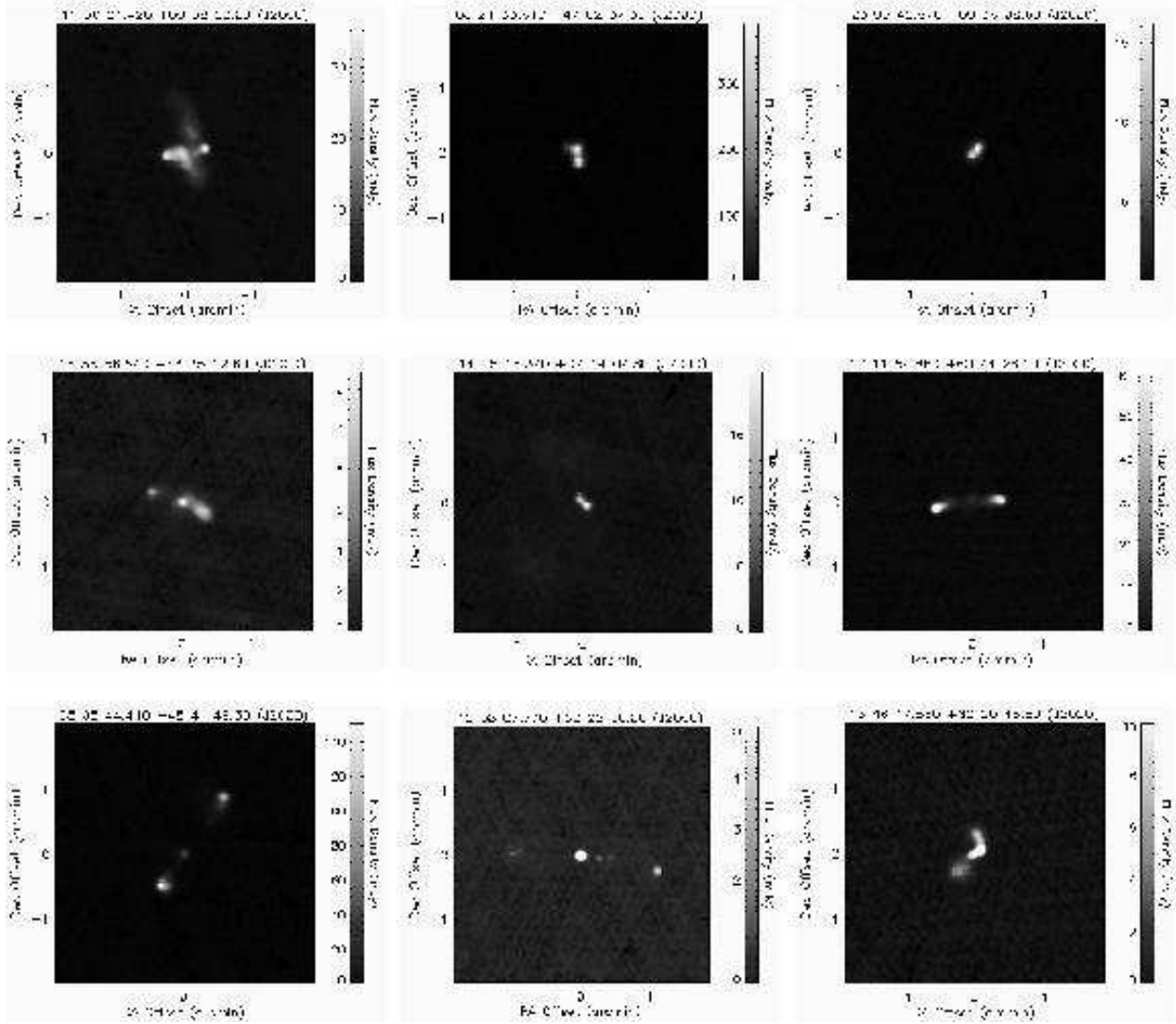}
\caption{Nine of the ten radio matches of double-peaked AGN with prominent 
  lobes. Note the peculiar radio morphology of SDSS J1130+0058 (top
  left, reminiscent of jet-reorientation) and the bow-like shape of
  SDSS J1346+6220 (bottom right).
\label{FIRSTlobes}} 
\end{figure}

The radio properties of disk-emission AGN described in the literature
place the majority of them in the radio-loud category, consistent with
the notion that disk-emission AGN are predominantly found in
low-accretion rate, large black hole mass, massive bulge, radio-loud
elliptical hosts \citep{EHsample94,ngc4450}. In fact, 16 of the 116
AGN in the main and auxiliary samples of disk-emitters were targeted
as FIRST sources (10/16 are also targeted as ROSAT sources, see
below), with all 16 also making the AGN target cut based on colors.
Since not all FIRST data were available at the time of SDSS object
targeting, we repeated the SDSS-FIRST match for the 102 of the 116
double-peaked AGN which currently fall in the area of overlap of the
two surveys. 31 sources out of these 102 have 20\,cm FIRST
counterparts within $2''$ of the SDSS position.  The $2''$ search
radius finds only core dominated sources and is smaller than the
distance at which the number of random SDSS-FIRST associations starts
becoming significant \citep[$2''.5$,][ hereafter I02]{sdssFirst}.
Visual inspection of $4'$\x$4'$ images from FIRST revealed 4
additional lobe dominated sources. There are a total of 10
lobe\footnote{One of the ten, SDSS J0229$-$0008, is a probable but not
  certain lobe dominated source. The tentative two lobes are 1$'$ and
  1.5$'$ away, respectively, and the closer one has a possible SDSS
  counterpart at $r\approx22$, about 6$''$ away from the radio
  position; thus it may be the chance superposition of three unrelated
  radio point sources.}  (see Fig.~\ref{FIRSTlobes}) and 25 core
dominated sources among the 35 FIRST detected objects. Two of the
extended radio sources have peculiar morphologies: SDSS J1130+0058
(top left in Fig.~\ref{FIRSTlobes}) is reminiscent of X-shaped jet
sources like 3C223.1 or 3C403, while SDSS J1346+6220 (bottom right in
Fig.~\ref{FIRSTlobes}) has a bent radio jet.

The main sample of 85 candidate disk-emitters match FIRST sources
within $60''$ (meant to include the lobe dominated sources as well, at
the expense of increased match-by-chance contamination) in \about 30\%
of the cases, compared to \about 19\% for the parent sample of 3216
AGN from which the selection of double-peaked profiles was made.
Limiting ourselves to the core dominated sources only (matched within
a $2''$ radius), \about 22\% of the disk-emission candidates are FIRST
sources, compared to \about 13\% for the general sample of AGN in the
same redshift range. In either case, candidate disk-emitters are
\about 1.6 times more likely to be radio sources than the average AGN
in the same redshift range. This difference is significant at the
$3\sigma$ level. The ratio of lobe to core dominated sources is about
2:5 for both the double-peaked and the parent AGN sample.

Table~\ref{tab5} lists the integrated 20\,cm luminosities for the 35
matched objects, using the sum of integrated flux densities to compute the
luminosity in the cases of lobe dominated objects. Following I02,
we define the ratio of radio to optical flux density, $R_i$ as:
\begin{equation}
R_i=\log(F\si{20\,cm}/F_i)=0.4(i-t)
\label{eqn0}
\end{equation}
where \iM\ is the \iM\ band SDSS PSF magnitude and $t$ is the 20\,cm
AB radio magnitude \citep{OkeGunn83} computed as
$t=-2.5\log{(F\si{20\,cm}/3631\textrm{Jy})}$ with $F\si{20\,cm}$ the
sum of integrated flux densities. Using the criterion $R_i>1$ for
radio-loudness, we find that 17 of the 35 objects with FIRST matches
are radio loud, 10 of which are in the auxiliary sample.  The
remaining 67 objects that are in the area covered by FIRST but not
detected, must have radio flux densities of less than the FIRST
detection limit of 1\,mJy and thus $t>16.4$. Since all but 6 of them
are brighter than $i=18.9$ in the optical, these additional 65
disk-emission candidates will have $R_i<0.4(18.9-16.4)=1.0$ and will
be radio-quiet. By using the PSF magnitudes here we minimize the host
galaxy contamination, so we do not expect this to have a large effect
on the radio-loudness estimation. According to this estimate \about
68\% of the disk emission candidates are radio-quiet. At most 37 of
all disk-emission candidates (32\%) could be radio loud, even if all
objects not found in the area covered by FIRST (14 objects) and all
undetected faint SDSS objects (6 objects with $i>18.9$) end up being
radio-loud. This is in stark contrast to the fact that almost all
previously known double-peaked AGN are radio-loud, as they were
selected from radio loud samples \citep{EHsample94}. If we repeat the
statistics using only the main sample, we find 7 out of the 85 are
detected in FIRST and radio-loud (8.2\%), 65$/$85 are radio-quiet
(76.5\%), and 13$/$85 (15.3\%) are too faint optically and not
detected in FIRST or not in FIRST coverage.

\begin{figure}
\epsscale{1.0}
\plotone{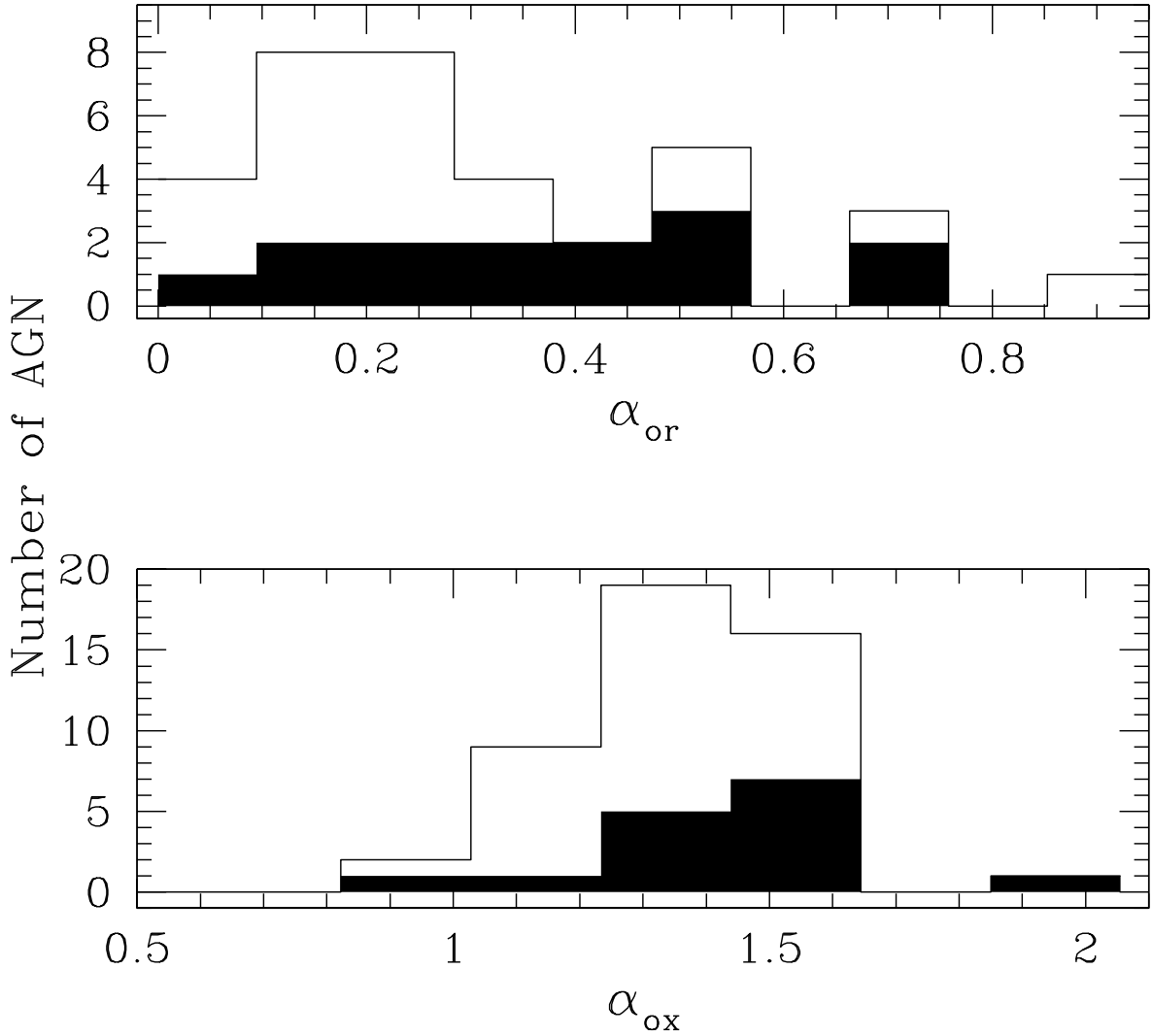}
\caption{Distribution of the optical-radio (top) and optical-X-ray
  (bottom) spectral indices for the FIRST and ROSAT detected AGN. The
  hollow histograms are for the main plus auxiliary sample, while the
  solid black histograms are for the auxiliary sample alone.
\label{specinx}} 
\end{figure}

The last column of Table~\ref{tab5} lists the optical to radio
spectral indices, with the optical flux K-corrected to that at
2500\,\AA\ (see eqn.~\ref{eqn2} below). Five of our disk-emission
candidates also match sources from the Westerbork Northern Sky Survey
\citep[WENSS,][]{WENSS} within 60$''$.  The WENSS is a long wavelength
($\lambda\lambda85,92$\,cm) radio survey of the sky north of
$\delta=30^{\circ}$ to a limiting flux density of \about 18\,mJy.  Three of
the FIRST-WENSS matched sources have radio lobes (SDSS J0806+4841,
SDSS J1638+4335, and SDSS J1238+5325) in the FIRST images and for two
of those (SDSS J0806+4841 and SDSS J1638+4335) the lobe emission
dominates the radio emission, resulting in different 20\,cm to 92\,cm
slopes of \hbox{$\alpha_{20}^{92}=-0.2$} and
\hbox{$\alpha_{20}^{92}=-1.1$}, respectively, compared to
\hbox{$\alpha_{20}^{92}=-0.5$} for the three core dominated sources.
We use the spectra index of the core dominated sources (more
appropriate for the majority of AGN), $\alpha_{20}^{92}=-0.5$,
assuming $F_{\nu} \propto \nu^{\alpha}$, to K-correct the FIRST flux
densities of all our sources to 20\,cm.  The optical to radio spectral
index is then computed as follows:
\begin{equation}
\alpha\si{or}=\left. - \log\left[{F_{\nu}(20\,\textrm{cm})
\over F_{\nu}(2500\,\textrm{\AA})}\right]\right/
\log{\left[\frac{2500\,\textrm{\AA}}{20\,\textrm{cm}}\right]}=0.1694
\log{\left[\frac{F_{\nu}(20\,\textrm{cm})}{F_{\nu}(2500\,\textrm{\AA})}\right]}
\label{eqn1}
\end{equation}
The distribution of $\alpha\si{or}$ indices is presented in the top
panel of Figure~\ref{specinx}.

\subsection{ROSAT counterparts}

For the general properties of a large sample of SDSS AGN matched with
ROSAT all-sky survey (RASS) catalogs we refer the reader to
\citet{Anderson03}.  In the case of our disk-emission sample, 45 of
the 116 AGN in the main and auxiliary samples were targeted as ROSAT
\citep{RASSb,RASSf} source counterparts for optical spectroscopy (all
but 5 of them were also targeted as AGN or galaxies based on their
optical photometry). We matched the 116 AGN from both the main and
auxiliary samples against the bright (18811 objects, search radius
$30''$, false detection rate 0.6\%) and faint source (105924 objects,
search radius $60''$, false detection rate 2.5\%) ROSAT all-sky
catalogs and found 47 ROSAT sources. Thus \about 41\% of all AGN in
our extended sample are detected in soft X-rays, \about 38\% for the
main sample and 48\% for the auxilary sample. For comparison, the
parent sample of 3216 AGN ($z<0.332$) have ROSAT matches in 28\% of
the cases. The candidate disk emitters from the main sample are thus
1.3 times more likely to have ROSAT counterparts.  This is at most
$2\sigma$ significant, considering that Poisson uncertainty alone
contributes 15\% to the error.  The higher fraction of detections in
the auxiliary sample is mainly due to its higher average redshift and
the corresponding selection of more high-luminosity AGN. We used
webPIMMs\footnote{http://heasarc.gsfc.nasa.gov/Tools/w3pimms.html} to
convert from count rate in the 0.1--2\,keV band to flux densities
using HI column densities from the Leiden-Dwingeloo 21\,cm maps
\citep{HImaps} to estimate the unobscured flux assuming a power law
continuum, $P(E)=E^{-\Gamma}$, with photon index $\Gamma=2$. For the
bright source catalog matches we verified that the hardness ratios are
consistent with a photon index of \about 2; the faint source hardness
ratios are too noisy for this purpose. The unobscured 0.1--2\,keV
luminosities, which are again relatively low compared to luminous AGN
(a few $\times 10^{44}$ erg\,s$^{-1}$ vs. $>10^{46}$ erg\,s$^{-1}$),
are given in Table~\ref{tab5}.

Using the \uM-\gM\ color, the \uM\ band (3543\,\AA) flux, and the
redshift $z$, we can estimate the flux density at 2500\,\AA:
\begin{equation}
F_{\nu}(2500\,\textrm{\AA})=3631\times10^{-0.4u-3.097\log[{1.417(1+z)}](u-g)}\:\textrm{Jy}
\label{eqn2}
\end{equation}
It is customary to define the optical to X-ray $\alpha\si{ox}$
spectral index as:
\begin{equation}
\alpha\si{ox}=\left . - \log\left[{F_{\nu}(2\,\textrm{keV}) \over F_{\nu}(2500\,\textrm{\AA})}\right]\right/
\log{\left[\frac{2500\,\textrm{\AA}}{2\,\textrm{keV}}\right]}=-0.3838
\log{\left[\frac{F_{\nu}(2\,\textrm{keV})}{F_{\nu}(2500\,\textrm{\AA})}\right]}
\label{eqn3}
\end{equation}
with the unobscured $F_{\nu}(2\,\textrm{keV})$ approximated by:
\begin{equation}
F_{\nu}(2\,\textrm{keV}) \approx \frac{\Delta F}{\Delta \nu}
=\frac{F(1.9 - 2\,\textrm{keV})\,\textrm{erg\,s}^{-1}\textrm{cm}^{-2}}{0.1\,\textrm{keV}\times4.836\times10^{17}\textrm{Hz\,keV}^{-1}}\times10^{23}\:\textrm{Jy}/(\textrm{erg\,s}^{-1}\textrm{Hz}^{-1}\textrm{cm}^{-2})
\label{eqn4}
\end{equation}
The optical to X-ray $\alpha\si{ox}$ spectral indices are given in the
ninth column of Table~\ref{tab5} and a histogram of the
$\alpha\si{ox}$ distribution is presented in the bottom panel of
Fig.~\ref{specinx}. Note that the auxiliary sample tends to have
larger spectral indices, but the majority of disk-emission candidates
have the canonical value of $\alpha\si{ox}=1.4$, indistinguishable from
that found by \citet{Anderson03} for the general sample of ROSAT-SDSS
matched AGN. 

This concludes our discussion of the observed properties of the
double-peaked AGN sample. We now proceed to compare the H$\alpha$ line
profiles to those of model accretion disks.
 
\section{Accretion Disk Models}

In this section we summarize the theoretical accretion disk models
used to simulate model emission line profiles. In the following
section we will use these to create a series of model lines, spanning
a range of disk parameters, for comparison with observations.

\subsection{Axisymmetric (Circular) Disk}

Prescriptions for computing model line profiles for circular accretion
disks were taken from \citet{circDisk}.  In brief, the models are
based on a simple relativistic Keplerian disk which is geometrically
thin and optically thick. Doppler boosting results in a higher blue
peak than red and a net redshift of the whole line is observed. In
order to reproduce the smoothness of the observed line profiles,
either a continuous emissivity law or local turbulent broadening is
added to the models; in what follows we use the local turbulent
broadening models. The specific intensity of the disk is:

\begin{equation}
I_{\nu_e}(\xi,\nu_e) =
\frac{\epsilon_0}{4\pi}\frac{\xi^{-q}}{\sqrt{2\pi}\sigma}
e^{\frac{(\nu_e-\nu_0)^2}{2\sigma^2}} = \frac{\epsilon_0}{4\pi}
\frac{\xi^{-q}}{\sqrt{2\pi}\sigma}
\exp\left[-\frac{(1+X-D)^2\nu_0^2}{2\sigma^2D^2}\right]
\label{eqn5}
\end{equation}

for $\xi_1 < \xi < \xi_2$, where $\xi \equiv r/R_G = rc^2/GM$ is the
dimensionless distance from the black hole in units in which the
gravitational constant and the speed of light are $G=c=1$, and $M$ is
the black hole mass and $R_G$ is the gravitational radius.  In
eqn.~\ref{eqn5}, $\nu_e$ is the emitted frequency, $\nu_0$ is the rest
frequency, $\nu$ is the observed frequency and $X \equiv \nu/\nu_0-1$.
The slope of the surface emissivity power law ($\epsilon_0\:\xi^{-q}$)
is $q$, while $\sigma/\nu_0$ is the dimensionless quantity
characterizing the local turbulent broadening.  The Doppler factor in
the weak field approximation is given by:

\begin{equation}
D = \sqrt{1-3/\xi}(1+\xi^{-1/2}\:\sin{i}\:\sin{\phi})^{-1}
\label{eqn6}
\end{equation}
where $i$ is the disk inclination ($90^\circ$ is edge-on, $0^\circ$ is
face-on), and $\phi$ is the azimuthal angle defined on the disk. Using
the Lorentz invariance of $I_{\nu_e}/\nu_e^3=I_{\nu}/\nu^3$, we obtain
an expression for the flux in the observer's frame:

\begin{equation}
F = \int\!\!\!\int\!\!\!\int I_{\nu_e}(\nu/\nu_e)^3\:d\nu\:d\Omega = \int
F_X\:dX
\label{eqn7}
\end{equation}
where $d\Omega$ is the solid angle subtended by the disk as seen by
the observer. If we denote the luminosity distance to the AGN as $d$,
the line profile becomes:

\begin{equation}
F_X = \frac{\epsilon M^2\nu_0\cos{i}}{d^2}\int_{\xi_1}^{\xi_2} \int_{-\pi/2}^{\pi/2} 
\xi\:I_{\nu_e}\:D^3\:g(D\:)d\phi\:d\xi
\label{eqn8}
\end{equation}
where
\begin{equation}
g(D) = 1 + \xi^{-1} \left[\frac{2 D^2}{D^2 \cos{i}^2 +
\xi[D-\sqrt{1-3/\xi}]^2} -1\right]
\label{eqn9}
\end{equation}
and the term of order $\xi^{-1}$ is the light bending correction.  To
obtain the observed spectrum in units of
erg\,cm$^{-2}$\,s$^{-1}$\,Hz$^{-1}$, one has to multiply F$_X$ from
eqn.~\ref{eqn8} above by $G^2/(\nu_0 c^4)$.

Computing a circular disk line profile amounts to performing the
double integration numerically in eqn.~\ref{eqn8} after specifying 
\emph{five parameters}: the inner and outer radii of the emitting
ring, $\xi_1$ and $\xi_2$, the disk inclination, $i$, the slope of the
surface emissivity power law, $q$, and the turbulent broadening
$\sigma$.  In theory, if these parameters were independent, they would
correspond to the five parameters of a two-Gaussian fit representation
of the double-peaked profile, resulting in a model that is fully
constrained by the data. While the line shape determines the five
model parameters given above, the overall normalization of the fit
defines the product of emissivity and black hole mass -- $\epsilon_0
M^2$ -- and this cannot lead to absolute black hole mass or emissivity
estimates since the models set the size of the emitting ring only in
relative units ({\it i.e.}, in gravitational radii).

\subsection{Elliptical Disk Model}

If the red peak is stronger than the blue, or the profile is observed
to be variable with successive blue and red dominant peaks
\citep{var}, the circular disk emission model fails and some asymmetry
in the disk must be invoked to reproduce the line asymmetry. Common
choices are elliptical disks (thought to arise when a single star is
disrupted in the vicinity of a black hole), warped disks
\citep[theorized to exist around rotating black
holes;][]{warped1,warped2}, spiral disks
\citep{spiral1,spiral2,spiral3} or disks with a hot spot
\citep{hotSpot}. Without extensive data on profile variability, these
models, which require many more free parameters (the elliptical disk
model adds two more parameters, the hot spot and the warped disk
models four extra parameters each), are often unjustified. In order to
compare our observed kinematic profiles to both axisymmetric and
non-axisymmetric disk models, we choose elliptical disk models to
represent all non-axisymmetric disks. This choice is justified only by
the relative simplicity of the computation, the addition of only two
more parameters and our inability to distinguish between the various
non-axisymmetric models with the current ``snapshot'' data.

In the case of an elliptical disk, the modified line profile
equations derived by \citet{ecDisk} are reproduced below. Two extra
parameters are added: the ellipticity, $e$, and the disk
orientation, $\phi_0$. The orientation angle is measured with respect
to the observer's line of sight, $\phi_0=0^\circ$ when the apocenter
points to the observer. The trajectory of particles in the disk is now
elliptical, so the distance from the black hole varies:
\begin{equation}
\xi(\phi) = \frac{\tilde{\xi}\:(1+e)}{1-e\:\cos{(\phi-\phi_0)}}
\label{eqn10}
\end{equation}

The line profile can again be calculated by integrating
eqn.~\ref{eqn8}, except $g(D)$ is replaced by $\Psi(\xi,\phi)$:
\begin{equation}
\Psi(\xi,\phi) = 1+\xi^{-1}\left(\frac{1-\sin{i}\cos{\phi}}{1+\sin{i}\cos{\phi}}\right)
\label{eqn11}
\end{equation}
The Doppler factor, $D$, in the elliptical disk case is:

\begin{equation}
\frac{1}{D}=\gamma\left\{\left(1-\frac{2}{\xi}\right)^{-1/2} - \nonumber \\
\frac{e\: \sin{(\phi-\phi_0)}\:\sqrt{1-(b/r)^2(1-2/\xi)}}
{\sqrt{\xi(1-2/\xi)^3\:(1-e\:\cos{(\phi-\phi_0)})}} +
\frac{\sin{i}\:\sin{\phi}\:(b/r)\:\sqrt{1-e\:\cos{(\phi-\phi_0)}}}
{\sqrt{\xi(1-2/\xi)(1-\sin^2{i}\cos^2{\phi})}} 
\right\}
\label{eqn12}
\end{equation}
where $(b/r)$ is the impact parameter in the weak field approximation:

\begin{equation}
\frac{b}{r} \approx \sqrt{1-\sin^2{i}\:\cos^2{\phi}}
\left[1+\frac{1}{\xi}\left(\frac{1-\sin{i}\cos{\phi}}{1+\sin{i}\cos{\phi}} \right) \right]
\label{eqn13}
\end{equation}
and $\gamma$ is the Lorentz factor:

\begin{equation}
\gamma = \left\{1-\frac{e^2\:\sin{(\phi-\phi_0)}+(1-2/\xi)[1-e\:\cos{(\phi-\phi_0)}]^2}
{\xi\:(1-2/\xi)^2[1-e\:\cos{(\phi-\phi_0)}]} \right\}^{-1/2}
\label{eqn14}
\end{equation}

\subsection{Grid of Disk Model Line Profiles}
\label{sec:modGrid}

\begin{figure}
\plotone{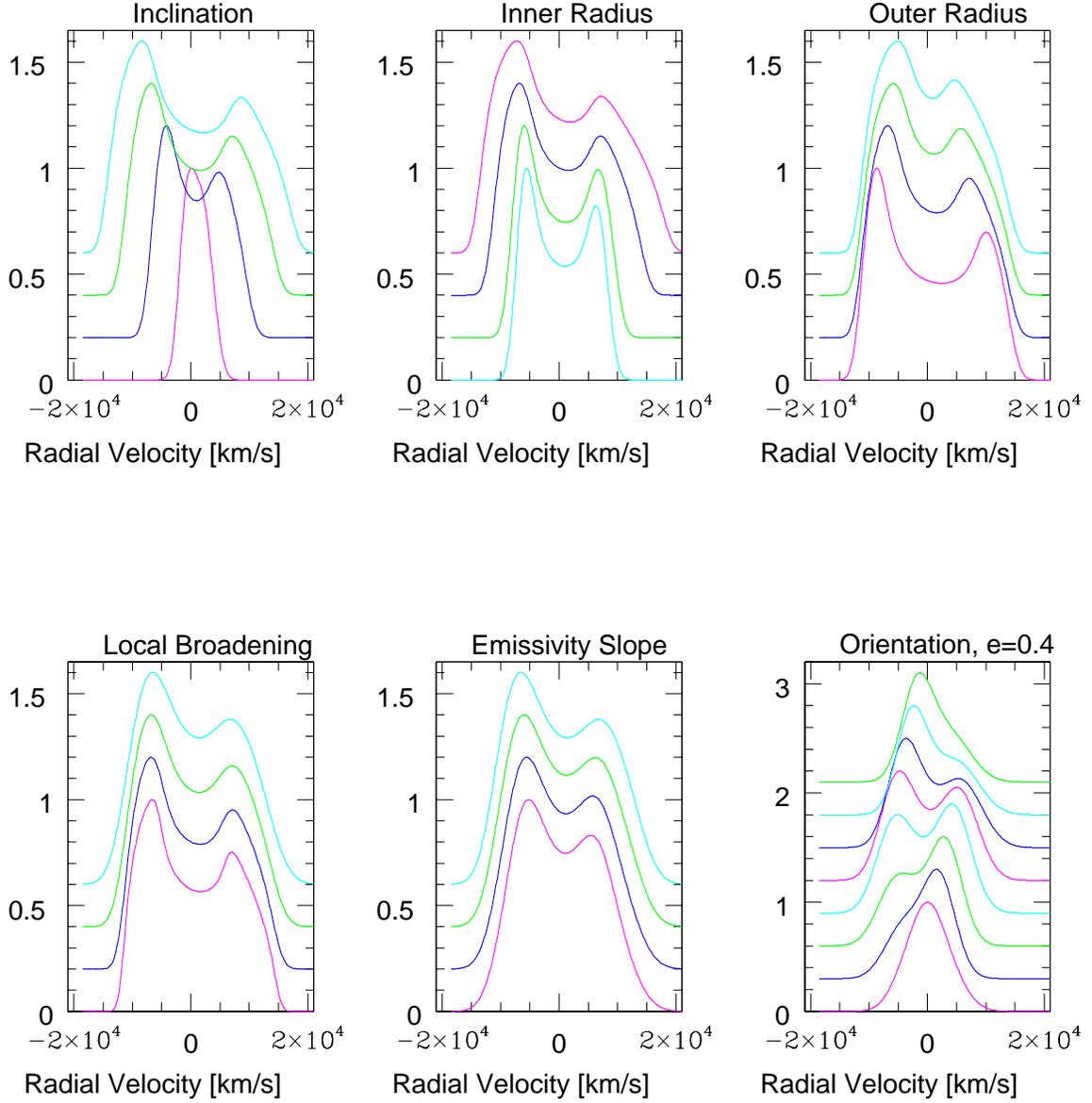}
\caption{\emph{Clockwise, from top left:} Variation of the line profile 
  with inclination $i$, inner radius $\xi_1$, outer radius $\xi_2$,
  local broadening $\sigma$, emissivity slope $q$, and elliptical disk
  orientation, $\phi_0$ ($e=0.4$). All lines shown are circular disk
  models, except the lower right panel. In each case the varying parameter
  \emph{increases} in value from the magenta to the cyan curve, taking
  all the values listed in Table~\ref{tab6}. The values of the
  remaining non-varying parameters in the order
  ($i$,$\xi_1$,$\xi_2$,$\sigma$,$q$,$\phi_0$,$e$) are (\emph{from top
    left}): ($...,3,300,1500,1200,0,0$), ($50,3,...,1500,1200,0,0$),
  ($50,3,300,...,1200,0,0$), ($50,3,300,1500,...,0,0$),
  ($50,...,300,1500,2400,0,0$), ($50,1.5,600,2500,2400,0.4,...$).
\label{grid}} 
\end{figure}

In order to understand the shapes of the line profiles and their
parameter dependence we created a grid of 24,000 elliptical disk
models (including the circular disk case for zero ellipticity), with
all possible combinations of the parameters listed in
Table~\ref{tab6}. The parameter ranges were chosen to cover the range
of known disk parameter estimates \citep{EHsample94}. Example model
line profiles are shown in Figure~\ref{grid}.

The effects of the disk model parameters on the computed line profiles
are as follows. Changing the disk inclination, $i$, has the most
dramatic effect on the line profile (see top left panel of
Fig.~\ref{grid}). Low inclination angles result in narrow observed
lines, which are frequently single peaked for $i<10^{\circ}$. The
inner radius\footnote{In what follows, we redefine $\xi$ to refer to
  radius in units of the gravitational radius, $R\si{G}$=GM/c$^2$.} of
the emitting ring, $\xi_1$, is smaller for broader line profiles and
the Doppler boosting of the blue relative to the red peak is more
pronounced for small $\xi_1$. The separation of the peaks is primarily
a function of the outer emitting disk radius, $\xi_2$.  As the outer
radius increases, the inclusion of slower moving material effectively
``fills in'' the gap between the peaks (top right panel of
Fig.~\ref{grid}), resulting in mostly single peak profiles for radii
of a few $\times 10^4 R_G$. The turbulent broadening, $\sigma$,
smooths the line profile, while slightly decreasing the blue to red
peak height ratio (bottom left panel of Fig.~\ref{grid}). For
radiation arising in local dissipation of gravitational energy in the
disk, the surface emissivity slope is expected to be $q=3$
\citep[similarly for a combination of direct radiation from an
elevated bulge and light scattered by a disk wind, the slope was found
to be in the range between 2.5 and 3,][]{MRW88}, but
\citet{EHsample94} find examples of different slopes in their fits to
observed profiles, some as small as $q=1.7$.  Decreasing the slope of
the surface emissivity makes the line profiles narrower, the blue side
of the lines less steep, and suppresses the blue peak height. For
non-axisymmetric disks, changing the orientation angle has a dramatic
effect on the observed line profile.  The lower right panel of
Figure~\ref{grid} shows 8 different orientations for a disk with
ellipticity $e=0.4$.  The profile changes from single to double peaked
and back, while the blue peak is not necessarily dominant in the
double-peaked phase. Note that these effects will be present also for
warped disks and disks with a hot spot, thus comparing elliptical disk
models to observed profiles will allow us to look for disk
asymmetries, whatever their nature.

\section{Comparison of Observed and Disk Model Line Parameters}
\label{sec:modComp}

Using the model line profiles, simulated over the parameter grid of
Table \ref{tab6}, we can estimate which accretion disk parameters are
consistent with the observed line shapes in a statistical sense.  We
chose this statistical comparison to individual accretion disk fits
for each disk-emission candidate because it's relatively
computationally inexpensive, does not constrain each AGN line-profile
to a specific model and can reveal the ranges of accretion disk model
parameters despite the degeneracies accompanying each profile
computation.

We perform the same line profile measurements on the model disk
profiles as on the observed H$\alpha$ lines: we measure the FWHM,
FWQM, their respective centroids, the positions of the blue and red
peaks and the blue-to-red peak-height ratios for all disk models that
actually show double peaks (a total of 784 models with $e=0$, and
13194 elliptical models).

\begin{figure}
\epsscale{1.0}
\plotone{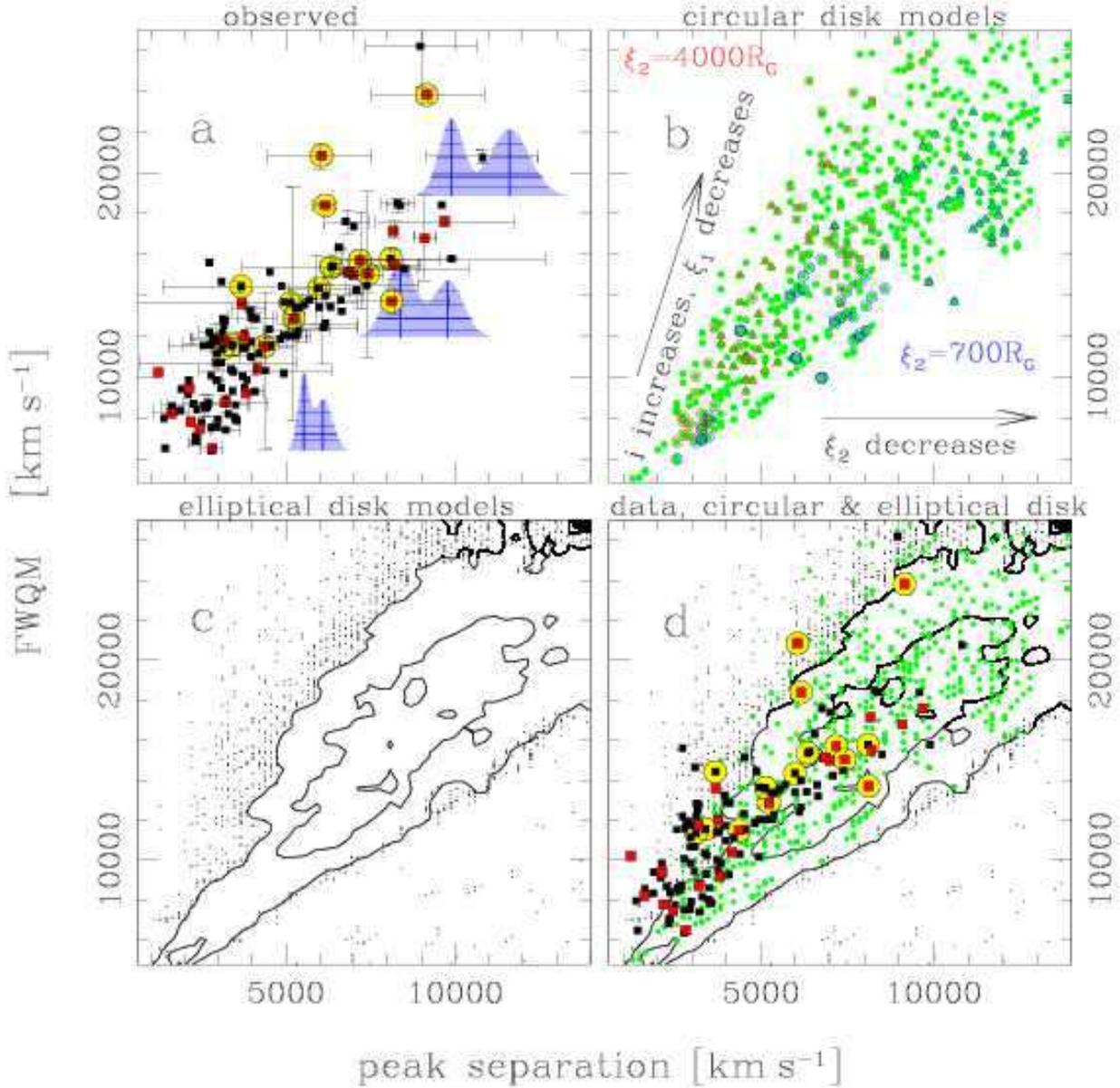}
\caption{Comparison of observed and model line quantities.
  a) Observed peak separation vs. FWQM and 3 Gaussian fits for
  illustration. The filled black squares are the main sample from
  Table~\ref{tab1}, the red squares are the auxiliary AGN from
  table~\ref {tab2}, and the large yellow circles denote radio-loud
  AGN; errorbars are given for the 63 AGN with alternative processing.
  b) Circular disk models. Each point corresponds to a double peak
  model realization. The red (blue) symbols correspond to models with
  largest (smallest) outer radii.  c) Non-axisymmetric (elliptical)
  disk models given as contours, the outliers as dots. d) Observed
  peak separation vs. FWQM compared to axisymmetric (green dots) and
  non-axisymmetric (black contours) disk models.
\label{modeldisk}}
\end{figure}

Figure~\ref{modeldisk} shows a comparison between observed and model
disk lines for two of the line measurements. The observed peak
separations vs.  the FWQM for the main and auxiliary samples are given
in panel a) with three example Gaussian fits for illustration.  As
noted above, the peak separation is a function of the outer radius,
$\xi_2$, while the FWQM is determined by both the inclination and the
inner radius of the emitting ring, $\xi_1$. The surface emissivity
slope, $q$, also affects the line width, which increases for larger q
values, but to a lesser extent. The top right panel gives the circular
disk models. The small outer radius ($\xi_2=700R_G$) models are
indicated in blue and the large outer radius ($\xi_2=4000R_G$) models
in red.  The elliptical disk models are presented in the lower left
panel with contours, and the lower right panel shows the observed line
parameters superimposed on both the elliptical and circular disk
models. This comparison suggests that the data prefer large outer
radii and disk inclinations smaller than 50$^{\circ}$, with both
axisymmetric and non-axisymmetric disk models agreeing with the data
in this projection.
 
\begin{figure}
\epsscale{1}
\plotone{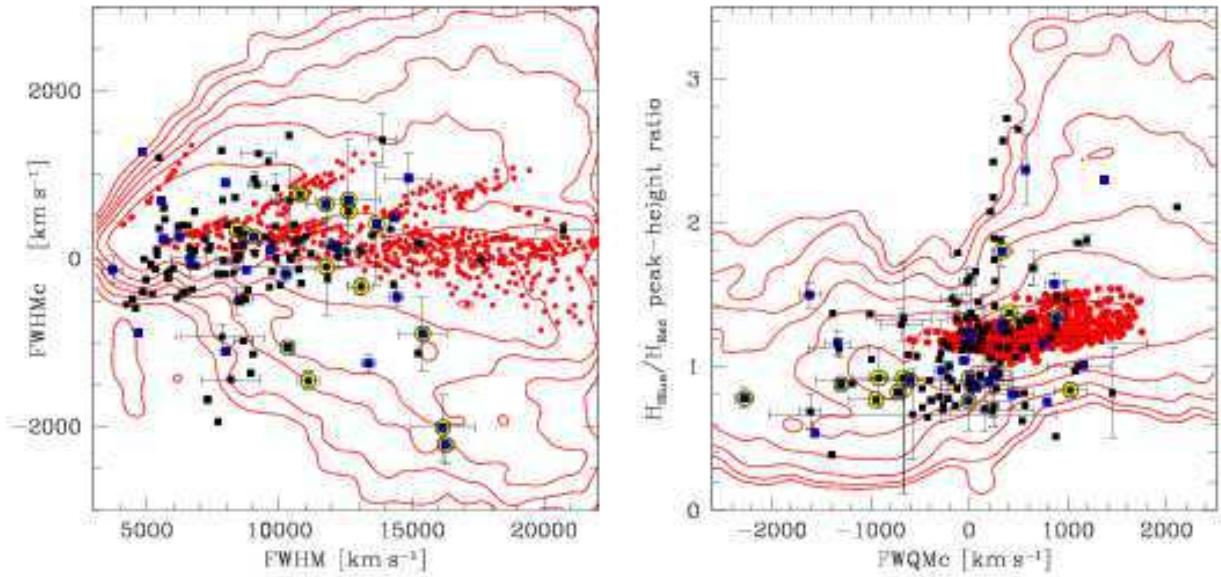}
\caption{Observables suggest the need for non-axisymmetric 
  disks. The observed values (black squares for the main sample, blue
  squares for the auxiliary sample, yellow circles around the
  radio-loud AGN) of the FWHM centroid ({\it left}) and the
  blue-to-red peak height ratio and FWQM centroid ({\it right}) are
  not all consistent with the circular disk models (red dots) in the
  sense that they can have FWHMc$<$0 and $H\si{blue}/H\si{red}<1$.
  The elliptical disk models (given as contours) are fully consistent
  with the data.
\label{asymdisk}} 
\end{figure}

Figure~\ref{asymdisk} presents another comparison of observed to model
line measurements, which suggests the need for non-axisymmetric disks
to explain the observed values of the centroids at FWHM (left panel)
and FWQM (right panel).  The observed line measurements (black squares
for the main sample, with errorbars from alternative processing,
whenever available) have a much larger range of values than do the
circular disk models (red dots), including negative centroids for the
FWQM of order $-$1000\,km\,s$^{-1}$ and stronger red peaks than blue
$H\si{blue}/H\si{red}<1$ in the right panel of Figure~\ref{asymdisk})
which are never realized for an axisymmetric disk. The elliptical disk
models (given as contours), however, appear to be able to account for
these line measurements.  This is a non-trivial statement, since there
are regions on the plot (e.g. the top and bottom left corners of the
left panel or the top left half of the right panel) which both the
models and the data avoid.

Using the seven line parameters (FWHM, FWQM, FWHMc, FWQMc,
$\lambda\si{red}$, $\lambda\si{blue}$, H$_{Blue}/$$H\si{red}$) we
compute the covariance matrix, $C$, for the observed line measurements
in the main sample. For any set of seven {\it model} line measurements
$\vec x$, we can compute the Gaussian equivalent probability that the
model line is consistent with the observed values: $P(x) = \exp
^{-1/2~\zeta^2}$, where $\zeta^2 = (\vec{x} - \left<\vec{x}\right>
)\cdot C^{-1} \cdot (\vec{x} - \left<\vec{x}\right> )$ is the squared
deviation for this model from the observed values. Given the inverse
of the covariance matrix (Table~\ref{tab7}) and the average vector
$\left<\vec{x}\right> = \{$$8852\textrm{\,km\,s}^{-1}$,
$11702\textrm{\,km\,s}^{-1}$, $-2\textrm{\,km\,s}^{-1}$,
$79\textrm{\,km\,s}^{-1}$, $2313\textrm{\,km\,s}^{-1}$,
$-1939\textrm{\,km\,s}^{-1}$, $1.2\}$ we select the model disks whose
measured line parameters are within $\zeta \leq 2$ of the observed
values. This procedure allows us to obtain the input disk parameters
(\emph{i.e.}  the inner and outer radii, the inclination, the surface
emissivity slope, and the turbulent broadening) without performing
detailed fits and constraining ourselves to a specific accretion disk
model, while keeping in mind the degeneracy in the model parameters
({\it e.g} the inner radius, inclination and surface emissivity
slope).

\begin{figure}
\epsscale{0.75}
\plotone{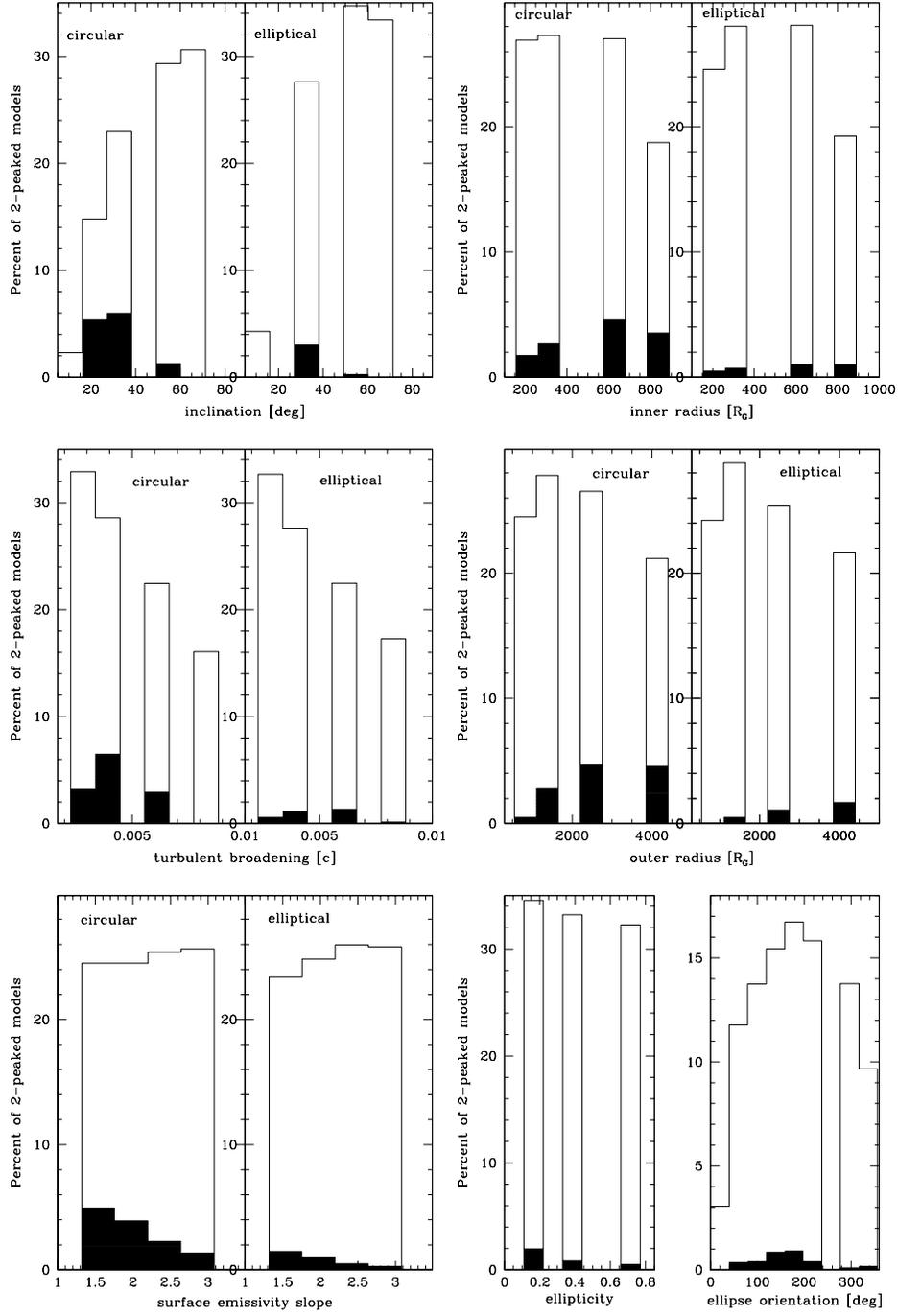}
\caption{Model parameters for disk models whose line profile 
  measurements are $\zeta<2$ away from the observed lines in the
  7-dimensional space of FWHM, FWQM, FWHMc, FWQMc, $\lambda\si{red}$,
  $\lambda\si{blue}$, and $H\si{blue}/H\si{red}$ are given in black shaded
  histograms (which sum up to a total of 12.6\% selected in the
  circular disk case and 3.4\% in the elliptical), compared to the
  initial range of model parameters resulting in double peaked lines
  (open histograms, sum up to 100\% for each of the circular and
  elliptical disk cases). Note that initially all model parameters
  were distributed equally in each direction, so a smaller percent of
  models within a particular open histogram bin means the rest were in
  single peaked lines.
\label{AllParam}} 
\end{figure}

The total number of model lines with double-peaked profiles in the
circular and elliptical disk cases are 784 and 13194, respectively,
out of a total of 1200 circular and 23040 elliptical disk model lines.
The $\zeta \leq 2$ criterion selects 99 out of 784 in the circular
disk case and 444 out of 13194 in the elliptical case.
Figure~\ref{AllParam} presents histograms of the model parameters for
the $\zeta \leq 2$ selected disks (shaded black) relative to the full
number of double-peaked models (open histograms). The percent
histograms in Figure~\ref{AllParam} are computed with respect to the
total number of double-peaked lines in both the circular and
elliptical disk case. In other words, the open histograms sum to 100\%
while the shaded ones to 12.6\% (99/784) in the circular and 3.4\%
(444/13194) in the elliptical disk case.

The initial set of disk models have equal numbers of models for each
parameter value (that is equal number of models were run, for example,
at each inclination). The fact that only 15\% of the circular disk
models shown as open histograms in Figure~\ref{AllParam} have
inclination of $i=20^{\circ}$ means that about 50\% of the
$i=20^{\circ}$ models are single peaked ($0.15\times784$ out of
$1200/5=240$ possible), compared to 90\% single peaked $i=10^{\circ}$
models and 25\% single peaked $i=30^{\circ}$ models for the circular
disk case. On the other hand, the circular and elliptical disk models
that were selected to have profiles within $\zeta \leq 2$ of the
observed data, have inclinations $20^{\circ}<i<30^{\circ}$ in \about
90\% of the cases.  That is, the data strongly favor smaller
inclination angles.  This result is in good agreement with the result
of \citet{Nandra97}, who argue that the Fe K$\alpha$ line profiles of
17 nearby ($z<0.05$) AGN are consistent with accretion disk emission
from disks with an average inclination of
$\left<i\right>=29^{\circ}\pm3$, with all but two cases suggesting
$i>50^{\circ}$.  The lack of disks with {\em very} small inclinations
($i<10^{\circ}$) to our line of sight is a selection effect, but there
is no a-priori reason why disks with high inclinations should be so
rare. A possible explanation is that for inclinations in excess of
$\about 50^{\circ}$ the obscuring torus, proposed to exist at larger
distances from the central engine, prevents direct observation of the
accretion disk, assuming that the small scale disk and large scale
torus are coplanar.  Whether they are coplanar is a matter of some
debate in the literature \citep[see the dust disk studies
of][]{Schmitt02,deKoff00}.  The inability of high inclination disk
models to match our data argues that there is in fact substantial
obscuration coplanar to the inner accretion disks of AGN.  If
larger-scale dust disks are not coplanar, perhaps the obscuration
inferred from our data arises from the outer accretion disks in the
form of a cool MHD disk wind \citep{ObsWind}.
 
The inclination $i$, inner radius $\xi_1$, and surface emissivity
slope $q$ are degenerate, in the sense that large $i$, small $\xi_1$
or large $q$ models give rise to line-profiles with large widths. As
long as the surface emissivity slope (for a surface emissivity defined
as $\epsilon_0\:\xi^{-q}$, $\xi_1<\xi<\xi_2$) is constrained to take
values between 1.5 and 3 (see below), the seven measured quantities
(FWHM, FWQM, FWHMc, FWQMc, $\lambda\si{red}$, $\lambda\si{blue}$,
H$_{Blue}$, and H$\si{red}$) are sufficient to isolate the best model
disk parameters consistent with the observations, despite the
degeneracy.  Due to relativistic effects in the strong gravity regime,
emission from regions close to the black hole not only increases the
line-profile width but also results in stronger line-asymmetries. For
example, a small $\xi_1$ will increase the dominance of the blue peak
together with the line width, while values of $q$ larger than 2 will
tend to make the blue wing less steep.

The observed line profiles prefer model disks with flatter surface
emissivity slopes ($1<q<2.5$, see bottom left panel of
Fig.~\ref{AllParam}) than the $q=3$ predicted for local gravitational
energy dissipation. In fact, if we allow $q$ to have even smaller
values ($0.5<q<1.5$), than the values found for disk-emission AGN in
the literature \citep[$1.5<q<3$,][]{EHsample94}, those would be
selected as consistent with the observed line-profiles, forcing about
a third of selected inclinations to values higher than $i=60^{\circ}$.
In other words, the preference of the data for small inclinations is
conditional on the surface emissivity having slopes steeper than
$q\approx2$. As mentioned in Section \ref{sec:modGrid}, all
theoretical models of disk illumination that we are aware of predict
surface emissivity decreasing with radius with a slope of about 3.
\citet{Dumont90} have developed a self consistent model for Balmer
line emission from geometrically thin, flaring Shakura-Sunyaev (1973)
disks. They find that disks (heated only by viscosity) tend to be too
cold at the radii corresponding to the Doppler shifts required by the
observed line-widths and argue that an illuminating source is
necessary to produce the Balmer line emission observed. The two disk
illumination models considered plausible by the authors --- a central
point source\footnote{One variant of the point source illumination
  model, which has a very large elevation of the radiation source
  above the disk, could result in a surface emissivity law which is
  almost flat with radius, if the disk emission comes from radii
  smaller than the source elevation.} and a diffuse medium that
backscatters the central source radiation --- both result in surface
emissivity slopes of 2--3. \citet{Rokaki92} use the Dumont and
Collin-Souffrin models to fit the Balmer line profiles together with
the full spectral energy distributions of six Seyfert galaxies (two of
which have double-peaked lines), and find that the diffuse medium
illumination model agrees well with the data, with surface emissivity
slopes between 2.1 and 3.6. In addition \citet{Nandra97} find that the
X-ray Fe K$\alpha$ line profiles of 17 nearby AGN are consistent with
accretion disk emission from the innermost ($6R_G<\xi<1000R_G$)
regions of the disk, with an average surface emissivity slope of
$\left<q\right>=2.5\pm0.4$, with all but one of the AGN preferring
$q>1.5$ (see their Figure 6). Taking into account the above mentioned
studies, we consider the high surface emissivity slope ($q>1.5$), low
inclination ($i<50^{\circ}$) models more likely.

The models selected to be within $\zeta \leq 2$ of the observed line
profiles in the main double-peaked sample tend to have larger outer
radii (typically $\xi_2>2000R_G$, see middle right panel of
Fig.~\ref{AllParam}), and local turbulent broadening of
$780$\,km\,s$^{-1}<\sigma<1800$\,km\,s$^{-1}$ (middle left panel of
Fig.~\ref{AllParam}). There is a slight preference for ellipticities
$e<0.4$ for non-axisymmetric models and no preference for the ellipse
orientation (bottom right panel of Fig.~\ref{AllParam}).

To quantify the effects seen in Figure~\ref{asymdisk} (which imply the
need for non-axisymmetric disk models) we compute the probability that
each of the observed triplets of FWHMc, FWQMc and blue-to-red peak
height ratio (H$_{Blue}/$$H\si{red}$) were drawn from the circular or
the elliptical disk pools of values. We compute new
$\zeta'$-ellipsoids\footnote{$\zeta'$ is again defined as $\zeta'^2 =
  \left(\vec{x'} - \left<\vec{x'}\right> \right)\cdot C^{-1} \cdot
  \left(\vec{x'} - \left<\vec{x'}\right> \right)$ , except here
  $\vec{x'}$ has only three components and the covariance matrix $C$
  is now a 3$\times$3 matrix.}  of both the circular and elliptical
model triplets and calculate for each observed triplet the number of
$\zeta'$ away from these average models.  In the circular disk case,
using the main observed sample, we find that 62\% of the observed
triplets have $\zeta'>3$ and are inconsistent the average axisymmetric
model. In contrast all but one of the 112 observations ($>99\%$) are
fully consistent with the elliptical models (\emph{i.e.} the observed
line measurements have $\zeta'<3$).

\section{Summary and Discussion} 

The present sample of double-peaked AGN, selected from the SDSS, is
the largest set of such objects to date, with 85 disk-emission
candidates in the main and an additional 31 candidates in an auxiliary
sample. The main sample was selected uniformly from all AGN observed
with SDSS with redshift $z<0.332$. The sample has H$\alpha$ lines
which are broader than those of the general AGN population
(FWHM$>5000$km\,s$^{-1}$), with larger red- and blueshifts of the
broad H$\alpha$ component, in agreement with previous disk-emission
sample statistics \citep{EHsample94}. The selected AGN are 1.6 times
more likely to be radio sources than the parent sample of $z<0.332$
AGN, but are predominantly (76\%) radio quiet. This is substantially
different from the existing sample of disk-emitters, which were mostly
sought (and found) in radio-loud subsamples of AGN.  The higher
fraction of double-peaked AGN found in radio-loud samples
\citep[\about 10\% for the radio-loud sample of ][compared to \about
3\% of all AGN with $z<0.332$ in this sample]{EHsample94} is probably
caused by the fact that both radio-loud quasars \citep[as opposed to
radio-loud galaxies,][]{McCarthy93} and double-peaked AGN tend to have
very broad emission lines and are rare.  If about 10\% of all AGN are
radio-loud and 10\% of them are disk-emitters we expect to find
double-peaked Balmer lines in only 1\% AGN of the general broad-line
AGN population.  We find about 3\%, the majority of them radio-quiet.
Previous studies have thus overlooked a large part of the disk
emission population consisting of radio-quiet AGN.

The selected double-peaked sample has medium luminosities, comparable
to those of the average broad line AGN at the same redshift (a few
$\times 10^{44}$erg\,s$^{-1}$). The equivalent widths of the low
ionization lines also do not differ significantly from those of the
majority of AGN, with the exception of [\ion{O}{1}]$\lambda$6300
equivalent widths and [\ion{O}{1}]/[\ion{O}{3}] flux ratios, which
tend to be larger for the double-peaked AGN. We find that \about 12\%
of the sample objects can be classified as LINERs according to
Kewley's selection criterion \citep{Kewley01}.

Comparison with accretion disk emission models suggest that about 60\%
of the lines must originate in non-axisymmetric disks. This is similar
to the fraction of double-peaked AGN that were not well fit by a
circular disk model found by \citet{EHsample94}.  We need long term
variability studies (the dynamical, thermal and sound-crossing times
of typical accretion disks at \about 1000$R_G$ are of order half a
year to 70 years for a 10$^8$\,M$_{\odot}$ black hole) to constrain
well the various disk model asymmetries that give rise to line profile
variation. A snapshot survey of the kind reported here cannot
distinguish between various disk asymmetries; this justifies the use
of one generic non-axisymmetric disk model (that of an elliptical
disk) to detect disk asymmetries.  The majority of radio-loud AGN are
highly variable; similarly the known disk-emitters which have been
followed up on timescales of decades (3C390.3, Arp\,102B, etc.) were
found to have line-profiles that varied significantly and required
asymmetric disk models.  Their identification with accretion disks
have been strengthened, not diminished, by variability studies. For
example, long term studies of double-peaked profile variations have
imposed a lower limit of the combined mass of a black hole binary in
excess of 10$^{10}$\,M$_{\odot}$, a result hard to reconcile with the
observed ranges of black hole masses \citep{BBBreject}.  Nonetheless,
it is possible that double-peaked line profiles have more than one
origin, and long term variability studies are essential to distinguish
between the various models. In view of this we have initiated a
follow-up variability study using Apache Point Observatory's 3.5\,m
telescope.  This should create, in time, a large sample with
variability information which will be better suited to reject or
confirm alternative models.

The binary black hole hypothesis for the double-peaked appearance of
the H$\alpha$ lines is quite tempting, in view of the fact that
galaxies merge and any black holes at their centers must pass through
a binary stage. One of our objects (SDSS J1130+0058, top left in
Figure \ref{FIRSTlobes}) has a radio morphology similar to radio-jet
reorientation objects like 3C52, 3C223.1, 3C403 and NGC 326. Radio-jet
reorientation can be caused by mergers of supermassive black holes
\citep{jetFlip2} or arise in the process of accretion disk realignment
if the black hole and disk axes are initially misaligned
\citep{jetFlip1}. In the case of a black hole merger, initially the
interaction of the black hole binary with the bulge stars will make
the orbit harder (i.e.  shrink the orbit), but beyond about 1\,pc, the
mechanism responsible for bringing the binary down to the scale where
gravitational radiation becomes important (10$^{-2}-10^{-3}$\,pc) is
not clear \citep{Merritt03,Yu02}. Hence no solid theoretical
predictions exist for the timescales associated with black hole
mergers, in particular the sub-pc regions of interest here.  Even if a
theory predicting supermassive black hole coalescence existed, the
fate of the gas feeding the activity adds more complexity.  As
\citet{Penston88} suggests, it is not obvious that the gas will stay
close to each black hole, forming two separate BLRs, especially for
unequal black hole mass mergers. It is plausible that the lower
velocity gas will orbit the center of mass of the system, resulting in
a single-peaked core, with two lower-intensity peaks emitted by the
faster moving material blended with the core to reveal a single peaked
profile \citep{CHF89}.

An interesting parallel between accreting compact objects in our
Galaxy (cataclysmic variables and low mass X-ray binaries) allows us
to speculate on the long term changes in AGN accretion disks by
analogy with the (much shorter timescale) evolution of the accretion
disks around the white dwarf or black hole primary in these systems.
\citet{CM97winds} propose a two zone line emission model for both AGN
and compact accreting objects. In their model, the high ionization
lines arise in the foot of the wind, just above the accretion disk
(where the radial component of the velocity is small but large radial
shear results in photons preferentially escaping from the back and
front of the disk); this gas is optically thick and gives rise to
single peaked lines. The low-ionization Balmer lines, on the other
hand, could originate in the optically thin atmosphere of the disk and
have double-peaked line profiles arising from disk rotation.

Dwarf novae, for example, are thought to be in a quiescent state, with
low accretion rates.  They show double peaked Balmer line profiles
(clearly originating in the disk as proved by secondary eclipses) and
have no high ionization lines. When they go into outburst, high
ionization, single-peaked lines appear in the disk wind. Nova-like
variables, in contrast, are almost constantly in outburst and very few
show double-peaked Balmer lines. They have high ionization lines (e.g.
\ion{He}{2}) and the high ionization lines are single-peaked. Chiang
and Murray's idea is that in outburst state (always for nova-like
variables, rarely for dwarf-novae), the accretion rate increases
dramatically and the radiation becomes powerful enough to drive a
wind. The high ionization (as well as the low ionization lines at
times) originate in the base (or sometimes higher up) in the disk
wind. In low, quiescent accretion states, the Balmer lines come from
the disk and are double-peaked, and there are no high ionization lines
coming from the disk.

Very few of the double-peaked AGN have been observed in the UV to date
with sufficient signal-to-noise ratio to consider high-ionization line
profiles (i.e.  \ion{C}{3}, \ion{C}{4}, Ly$\alpha$). The prototype
disk-emitter, Arp\,102B, has single peaked high-ionization lines
narrower than the Balmer lines \citep{uvspec}.  If this fact remains
true for the majority of AGN with double-peaked Balmer lines, we could
draw a direct parallel between the double-peaked AGN and quiescent
dwarf-novae. Current findings classify double-peaked AGN as
predominantly low-luminosity, low accretion rate systems relative to
the majority of AGN \citep{EHsample94,ngc4450}.  The lack of big blue
bump, combined with new theoretical models of low-rate accretion, 
suggests that the inner part of the disk is elevated into the
quasi-spherical structures \citep[the ``ion torus'' of][]{CHF89} of
Advection Dominated Accretion Flow (ADAF) or Advection Dominated
Inflow-Outflow (ADIOS) models \citep{IAN00,IN02,BNQ01}. Thus AGN with
double-peaked Balmer lines could be viewed as the supermassive black
hole analogs of dwarf-novae, in states of low accretion and
luminosity, with the broad low-ionization emission originating in the
parts of the disk just outside the elevated ion-torus, providing the
excess energy and high-ionization lines arising in a weak wind.
According to this view, the accretion rate and radiation pressure in
the majority of AGN is substantially higher, resulting in a radiation
driven \citep[or MHD driven, see][]{Proga02} wind and predominantly
single peaked high and low ionization lines.

\vskip 0.5in \leftline{Acknowledgments} We are grateful to Michael
Eracleous, Luis Ho, Jules Halpern, Bogdan Paczynski, and Wei Zheng for
their advice and stimulating discussions; we also acknowledge the
helpful comments of an anonymous referee. We wish to thank Doug
Finkbeiner for his help with obtaining the HI column densities from
the Leiden-Dwingeloo 21\,cm maps. IVS and MAS acknowledge the support
of NSF grant AST-0071091. IVS acknowledges the NASA Applied
Information Systems Research Program (AIRSP) for partial support.
Funding for the creation and distribution of the SDSS Archive has been
provided by the Alfred P.  Sloan Foundation, the Participating
Institutions, the National Aeronautics and Space Administration, the
National Science Foundation, the U.S. Department of Energy, the
Japanese Monbukagakusho, and the Max Planck Society.  The SDSS Web
site is http://www.sdss.org/.  The SDSS is managed by the
Astrophysical Research Consortium (ARC) for the Participating
Institutions. The Participating Institutions are The University of
Chicago, Fermilab, the Institute for Advanced Study, the Japan
Participation Group, The Johns Hopkins University, Los Alamos National
Laboratory, the Max-Planck-Institute for Astronomy (MPIA), the
Max-Planck-Institute for Astrophysics (MPA), New Mexico State
University, University of Pittsburgh, Princeton University, the United
States Naval Observatory, and the University of Washington.

\appendix
\section{The Auxiliary Sample of Double-peaked AGN}
\label{sec:Auxiliary}

Thirty-one objects of interest were not identified by one of the two
steps of the selection algorithm but have line profiles suggestive of
disk emission and are included in this auxiliary sample.  The list of
additional objects was presented in Table~\ref{tab2}. The first column
lists the object name in the format ``SDSS Jhhmmss.s$\pm$ddmmss.s'',
J2000, the second the redshift, columns three through seven give the
apparent model magnitude corrected for Galactic extinction, and the
last column contains selection comments.

The majority (21) of the 31 additional objects listed in
Table~\ref{tab2} were missed on the PCA step, because their redshift
was too high ($z>0.332$, 14 objects marked ``HiZ'', including the
three $z\approx0.6$ objects with \ion{Mg}{2} selection), because they
were flagged as having spectral defects (two objects marked
``badFlag''), because the automatic redshift estimate failed to
recognize the H$\alpha$ line and assigned an incorrect high redshift
(four objects marked ``wrongZ'') or because they failed the
eigen-spectra coefficient cut (one object marked ``NotPCASel''). All
but five of those 21 were fit with multiple Gaussians and selected by
the double-peak finding step. The five AGN missed on the PCA step for
which only part of the H$\alpha$ line is included in the spectral
coverage (2/5) or H$\alpha$ is out of range (3/5) are flagged
``NoParam'' in Table~\ref{tab2}.

The remaining ten auxiliary objects listed in Table~\ref{tab2} (marked
``NotGausSel'') were missed during the Gaussian fit selection because
the two peaks are blended and were missed by the inflection point
selection (8/10) or because of a wrong set of initial parameters in a
many-Gaussian fit (2/10). Eight of the ten were run through the peak
finding algorithm with relaxed inflection point thresholds (the
remaining 2 are flagged ``NoParam'' in Table~\ref{tab2}).

\section{Vacuum Wavelengths}
\label{sec:vacWave}

The wavelengths in SDSS spectra are given in vacuum, corrected to the
heliocentric reference frame. Given a vacuum wavelength
$\lambda_{vac}$ in \AA, the conversion to air wavelength is given by
\citep{Morton91}:
\begin{equation}
\lambda_{air}=\frac{\lambda_{vac}}{(1.0+2.735182\times10^{-4}+131.4182/\lambda_{vac}^2+2.76249\times10^8/\lambda_{vac}^4)}
\end{equation}
The conventional air wavelengths and the SDSS vacuum wavelengths for the lines
commonly used in this paper are listed in Table \ref{tab8}.

\clearpage

\clearpage
 
\clearpage

\clearpage 

\clearpage 
 
\clearpage 

\clearpage 
 
\clearpage 

\clearpage 
 
\clearpage 

\clearpage 

\clearpage 

\clearpage 

\clearpage 

\clearpage 

\clearpage 

\clearpage 

\clearpage 

\clearpage 

\clearpage 

\clearpage 
\begin{deluxetable}{cccccccc} 
  \tabletypesize{\footnotesize} \tablenum{1} \tablecaption{Double Peak
    H$\alpha$ AGN sample: Coordinates and Apparent Magnitudes}
  \tablehead{\colhead{SDSS ID} & \colhead{Redshift} &\colhead{$u$}
    &\colhead{$g$} &\colhead{$r$} &\colhead{$i$} &\colhead{$z$}
    &\colhead{Comments}} \startdata
  SDSS J000710.02+005329.0 & 0.3159 & 17.04 & 16.76 & 16.56 & 16.64 & 16.01 & 2P+C, repeat1,337   \\
  SDSS J000815.46$-$104620.5 & 0.1986 & 19.23 & 18.06 & 17.12 & 16.61 & 16.42 & RS   \\
  SDSS J001224.03$-$102226.3 & 0.2200 & 17.07 & 17.03 & 16.83 & 16.49 & 16.62 & RS   \\
  SDSS J004319.75+005115.3 & 0.3083 & 17.84 & 17.91 & 17.80 & 17.75 & 16.93 & 2P+C, repeat1,332   \\
  SDSS J005709.93+144610.3 & 0.1718 & 16.02 & 16.01 & 15.82 & 15.36 & 15.58 & RS+C   \\
  SDSS J011140.03$-$095834.9 & 0.2064 & 19.08 & 18.56 & 17.77 & 17.32 & 17.04 & 2B   \\
  SDSS J013253.31$-$095239.5 & 0.2597 & 18.42 & 18.37 & 18.10 & 17.88 & 17.49 & 2P, repeat1,31   \\
  SDSS J013407.89$-$084129.9 & 0.0699 & 19.07 & 17.72 & 17.02 & 16.56 & 16.32 & 2B, repeat1,31   \\
  SDSS J014901.09$-$080838.1 & 0.2093 & 19.73 & 19.25 & 18.44 & 17.94 & 17.79 & RS   \\
  SDSS J022014.57$-$072859.2 & 0.2136 & 18.22 & 18.24 & 17.63 & 17.04 & 16.98 & MG   \\
  SDSS J023253.42$-$082832.1 & 0.2652 & 17.98 & 17.82 & 17.26 & 16.94 & 16.55 & RS   \\
  SDSS J024052.82$-$004110.9 & 0.2466 & 18.59 & 18.14 & 17.50 & 17.28 & 17.15 & 2P+C, repeat1,378   \\
  SDSS J024703.24$-$071421.7 & 0.3340 & 19.89 & 19.60 & 18.76 & 18.40 & 17.55 & RS+C, RL   \\
  SDSS J024840.04$-$010032.7 & 0.1840 & 19.44 & 18.38 & 17.49 & 16.97 & 16.64 & 2P   \\
  SDSS J025220.89+004331.3 & 0.1696 & 17.93 & 17.70 & 17.21 & 16.81 & 16.61 & 2B, repeat1,61   \\
  SDSS J025951.73$-$001522.7 & 0.1018 & 18.89 & 18.31 & 17.87 & 17.30 & 17.31 & RS, repeat4,30,358   \\
  SDSS J034931.03$-$062621.0 & 0.2877 & 18.35 & 18.29 & 17.93 & 17.72 & 17.31 & RS+C   \\
  SDSS J073927.24+404347.4 & 0.2081 & 19.58 & 18.96 & 18.48 & 18.13 & 18.07 & 2P   \\
  SDSS J074157.26+275519.9 & 0.3256 & 18.82 & 18.46 & 18.13 & 18.09 & 17.40 & 2B   \\
  SDSS J075930.77+352803.8 & 0.2886 & 20.54 & 20.31 & 19.62 & 19.32 & 18.88 & RS+C   \\
  SDSS J080310.58+293233.8 & 0.3277 & 18.99 & 18.98 & 18.68 & 18.64 & 17.89 & 2P+C   \\
  SDSS J081329.29+483427.9 & 0.2738 & 18.05 & 18.04 & 17.87 & 17.87 & 17.38 & 2P, repeat1,27  \\
  SDSS J081700.40+343556.3 & 0.0620 & 18.95 & 17.36 & 16.42 & 15.73 & 15.56 & 2P+C, RL   \\
  SDSS J081916.30+481745.5 & 0.2228 & 18.53 & 18.60 & 18.05 & 17.52 & 17.38 & RS, repeat1,27   \\
  SDSS J082113.71+350305.0 & 0.2936 & 18.82 & 18.52 & 18.17 & 18.07 & 17.73 & BS   \\
  SDSS J082125.37+421908.5 & 0.2220 & 19.06 & 18.77 & 18.22 & 17.77 & 17.66 & RS   \\
  SDSS J082133.61+470237.3 & 0.1283 & 18.93 & 17.56 & 16.67 & 16.14 & 15.94 & RS, RL   \\
  SDSS J082205.25+455349.2 & 0.2998 & 18.11 & 18.02 & 17.64 & 17.52 & 17.08 & 2B   \\
  SDSS J082406.23+334244.9 & 0.3179 & 19.02 & 18.81 & 18.26 & 18.33 & 17.51 & 2P   \\
  SDSS J083225.35+370736.2 & 0.0920 & 15.91 & 15.96 & 16.05 & 15.45 & 15.73 & MG+C   \\
  SDSS J083826.50+371906.7 & 0.2110 & 19.27 & 19.10 & 18.52 & 17.99 & 18.01 & RS   \\
  SDSS J084110.88+022952.1 & 0.3322 & 18.88 & 18.96 & 18.76 & 18.69 & 18.03 & BS   \\
  SDSS J084535.38+001619.5 & 0.2613 & 19.31 & 18.96 & 18.34 & 17.96 & 17.52 & 2P   \\
  SDSS J091459.05+012631.3 & 0.1977 & 18.28 & 17.63 & 16.80 & 16.38 & 16.17 & 2P+C, repeat1,26   \\
  SDSS J091828.60+513932.1 & 0.1855 & 17.63 & 17.34 & 17.07 & 16.71 & 16.72 & 2B   \\
  SDSS J092515.00+531711.8 & 0.1862 & 19.22 & 17.84 & 16.84 & 16.31 & 16.09 & RS+C   \\
  SDSS J093509.47+481910.2 & 0.2237 & 18.16 & 17.79 & 17.42 & 16.99 & 17.27 & MG   \\
  SDSS J093844.46+005715.8 & 0.1704 & 18.50 & 17.30 & 16.47 & 15.50 & 15.65 & RS, repeat1,287   \\
  SDSS J100443.44+480156.5 & 0.1986 & 19.45 & 18.67 & 17.89 & 17.36 & 17.29 & 2P+C   \\
  SDSS J101405.89+000620.3 & 0.1412 & 19.38 & 17.79 & 16.81 & 16.21 & 15.95 & 2P   \\
  SDSS J103202.40+600834.5 & 0.2939 & 19.26 & 19.01 & 18.12 & 17.70 & 17.23 & RS   \\
  SDSS J104108.19+562000.4 & 0.2304 & 19.70 & 19.32 & 18.27 & 17.92 & 17.83 & 2P+C   \\
  SDSS J104128.60+023205.0 & 0.1820 & 19.25 & 18.71 & 17.89 & 17.35 & 17.13 & 2P   \\
  SDSS J104132.78$-$005057.5 & 0.3029 & 18.11 & 18.00 & 17.71 & 17.59 & 16.98 & MG+C   \\
  SDSS J110742.77+042134.2 & 0.3269 & 18.87 & 18.82 & 18.36 & 18.08 & 17.55 & BS, repeat1,3   \\
  SDSS J112751.95+675042.8 & 0.1940 & 18.17 & 18.02 & 17.83 & 17.44 & 17.51 & RS   \\
  SDSS J113021.00+022211.5 & 0.2410 & 17.48 & 17.54 & 17.48 & 17.16 & 17.20 & BS+C   \\
  SDSS J113021.42+005823.0 & 0.1325 & 17.98 & 16.79 & 15.88 & 15.46 & 15.25 & BS, RL   \\
  SDSS J113633.09+020747.5 & 0.2390 & 18.35 & 18.00 & 17.38 & 16.90 & 16.74 & 2B   \\
  SDSS J114335.36$-$002942.4 & 0.1715 & 18.01 & 17.82 & 17.30 & 16.77 & 16.83 & BS+C, repeat2,76,375   \\
  SDSS J115047.48$-$031652.9 & 0.1486 & 18.98 & 17.81 & 17.06 & 16.64 & 16.34 & 2P   \\
  SDSS J115227.12+604817.5 & 0.2703 & 19.39 & 19.26 & 18.58 & 18.32 & 17.77 & RS+C   \\
  SDSS J115644.11+614741.6 & 0.2265 & 17.88 & 17.80 & 17.60 & 17.19 & 17.16 & RS+C   \\
  SDSS J121855.80+020002.2 & 0.3270 & 17.43 & 17.26 & 17.13 & 17.26 & 16.55 & BS+C   \\
  SDSS J122009.55$-$013201.2 & 0.2879 & 18.57 & 18.58 & 18.28 & 18.10 & 17.56 & 2B   \\
  SDSS J130927.67+032251.8 & 0.2665 & 18.50 & 18.60 & 18.37 & 18.17 & 17.73 & 2B+C, repeat1,266  \\
  SDSS J132834.14$-$012917.6 & 0.1515 & 18.36 & 18.05 & 17.45 & 16.80 & 16.85 & BS, RL   \\
  SDSS J133312.43+013023.7 & 0.2171 & 19.00 & 18.78 & 18.30 & 17.84 & 17.93 & RS, RL   \\
  SDSS J135107.04+653127.4 & 0.2988 & 18.47 & 18.49 & 18.33 & 18.37 & 17.67 & BS+C   \\
  SDSS J140019.28+631427.0 & 0.3309 & 18.24 & 17.96 & 17.57 & 17.47 & 17.10 & 2P   \\
  SDSS J140720.70+023553.1 & 0.3094 & 19.57 & 19.41 & 18.73 & 18.45 & 17.80 & 2P+C   \\
  SDSS J141454.55+013358.6 & 0.2704 & 19.52 & 19.53 & 18.71 & 18.35 & 17.81 & BS   \\
  SDSS J141613.37+021907.8 & 0.1580 & 18.90 & 17.84 & 16.84 & 16.27 & 16.13 & RS+C, RL   \\
  SDSS J141946.04+650353.0 & 0.1478 & 18.38 & 17.75 & 17.05 & 16.58 & 16.43 & BS, repeat1,4   \\
  SDSS J142424.22+595300.6 & 0.1348 & 15.95 & 15.92 & 15.83 & 15.51 & 15.56 & RS   \\
  SDSS J142754.77+635448.4 & 0.1453 & 18.36 & 17.86 & 17.18 & 16.68 & 16.48 & 2P   \\
  SDSS J143455.31+572345.3 & 0.1749 & 17.32 & 17.06 & 16.87 & 16.45 & 16.42 & BS\tablenotemark{1}, repeat2,8,95  \\
  SDSS J154019.58$-$020505.4 & 0.3200 & 16.39 & 16.39 & 16.33 & 16.42 & 15.88 & 2P+C   \\
  SDSS J154534.55+573625.1 & 0.2681 & 19.17 & 18.75 & 18.00 & 17.73 & 17.29 & 2B   \\
  SDSS J160548.03$-$010913.0 & 0.2425 & 19.73 & 19.28 & 18.27 & 17.79 & 17.56 & RS   \\
  SDSS J170102.29+340400.6 & 0.0945 & 18.56 & 18.05 & 17.59 & 16.67 & 16.90 & BS+C, repeat1,7   \\
  SDSS J171806.84+593313.4 & 0.2728 & 18.74 & 18.89 & 18.61 & 18.43 & 17.96 & 2P   \\
  SDSS J172102.47+534447.2 & 0.1918 & 19.10 & 18.76 & 18.12 & 17.60 & 17.51 & BS+C   \\
  SDSS J172711.83+632241.9 & 0.2175 & 17.12 & 17.04 & 16.93 & 16.66 & 16.74 & 2P, repeat1,95  \\
  SDSS J173038.27+550016.7 & 0.2491 & 19.25 & 18.98 & 18.54 & 18.17 & 17.91 & 2B, repeat4,3,219   \\
  SDSS J210109.57$-$054747.3 & 0.1794 & 18.23 & 17.91 & 17.15 & 16.66 & 16.41 & 2P   \\
  SDSS J211353.33$-$061241.2 & 0.2411 & 19.48 & 19.25 & 18.91 & 18.62 & 18.59 & RS   \\
  SDSS J214555.04+121034.2 & 0.1113 & 18.52 & 18.01 & 17.46 & 16.96 & 16.81 & 2B   \\
  SDSS J214935.23+113842.1 & 0.2393 & 20.13 & 19.71 & 18.84 & 18.41 & 18.16 & 2P+C   \\
  SDSS J222913.62+000840.9 & 0.2657 & 19.07 & 18.32 & 17.34 & 16.93 & 16.51 & 2P   \\
  SDSS J223336.71$-$074337.1 & 0.1750 & 19.14 & 18.55 & 17.76 & 17.20 & 16.99 & BS   \\
  SDSS J230443.47$-$084108.6 & 0.0471 & 15.69 & 15.27 & 15.08 & 14.59 & 14.52 & BS   \\
  SDSS J231254.90$-$011620.6 & 0.2139 & 18.74 & 18.38 & 17.88 & 17.62 & 17.38 & 2P   \\
  SDSS J232721.96+152437.3 & 0.0460 & 16.22 & 15.11 & 14.44 & 13.95 & 13.64 & BS+C   \\
  SDSS J233254.46+151305.4 & 0.2146 & 17.53 & 17.55 & 17.34 & 16.99 & 17.02 & 2P 
  \enddata 
  \tablecomments{The apparent magnitudes quoted here (\emph{Photo}
    version 5.3) are total {\it model} galaxy magnitudes computed by
    convolving an exponential or de Vaucouleurs model with the PSF.
    The uncertainty in the magnitudes is at the few percent level,
    mostly due to photometric calibration, except in a few cases for
    \uM~and \zM~band measurements, where photon statistics dominate
    the error.}
\tablenotetext{1}{Not corrected for telluric absorption 7600-7700\AA.}
\label{tab1}\end{deluxetable}

\clearpage 
\begin{deluxetable}{cccccccc} 
\tablewidth{0pt}
\tabletypesize{\footnotesize}
\tablenum{2}
\tablecaption{Double Peak H$\alpha$ Auxiliary Sample: Coordinates and Apparent Magnitudes} 
\tablehead{\colhead{SDSS ID} & \colhead{Redshift}
&\colhead{$u$} &\colhead{$g$} &\colhead{$r$}
&\colhead{$i$} &\colhead{$z$} &\colhead{Comments}} 
\startdata
SDSS J011734.84$-$011135.6 & 0.1855 & 19.23 & 18.68 & 17.89 & 17.31 & 17.33 & NotGausSel \\
SDSS J021259.60$-$003029.6 & 0.3942 & 17.71 & 17.53 & 17.62 & 17.44 & 17.06 & HiZ, NoParam \\
SDSS J021655.88$-$005228.9 & 0.2778 & 19.82 & 19.70 & 18.94 & 18.86 & 18.27 & NotGausSel \\
SDSS J022930.92$-$000845.4 & 0.6091 & 20.11 & 19.66 & 19.51 & 19.00 & 18.94 & HiZ, NoParam, \ion{Mg}{2}Select, RL \\
SDSS J030021.41$-$071458.9 & 0.3883 & 16.88 & 16.63 & 16.82 & 16.81 & 16.55 & HiZ, NoParam \\
SDSS J032559.97+000800.8 & 0.3602 & 17.08 & 17.06 & 17.13 & 17.18 & 16.71 & HiZ, repeat1,32 \\
SDSS J075407.96+431610.6 & 0.3475 & 17.28 & 17.18 & 17.13 & 17.19 & 16.57 & HiZ, RL \\
SDSS J080644.41+484149.2 & 0.3700 & 17.51 & 17.48 & 17.52 & 17.49 & 17.00 & HiZ, repeat1,27, RL \\
SDSS J090436.96+553602.7 & 0.0372 & 16.89 & 16.55 & 16.16 & 15.87 & 15.68 & MG NotGausSel, NoParam\\
SDSS J093653.85+533126.9 & 0.2281 & 17.29 & 17.06 & 16.86 & 16.47 & 16.60 & NotGausSel \\
SDSS J100027.44+025951.3 & 0.3390 & 19.10 & 18.50 & 17.99 & 17.86 & 17.21 & HiZ \\
SDSS J102738.54+605016.5 & 0.3314 & 17.75 & 17.51 & 17.36 & 17.28 & 16.92 & wrongZ, RL \\
SDSS J114051.59+054631.1 & 0.1315 & 19.58 & 17.91 & 16.99 & 16.50 & 16.15 & wrongZ \\
SDSS J121154.86+604426.1 & 0.6370 & 20.02 & 19.63 & 19.70 & 19.37 & 19.32 & HiZ, NoParam, \ion{Mg}{2}Select, RL \\
SDSS J123807.77+532556.0 & 0.3478 & 17.26 & 17.30 & 17.30 & 17.33 & 16.69 & HiZ, RL \\
SDSS J132442.44+052438.8 & 0.1154 & 18.81 & 18.58 & 18.36 & 17.53 & 17.82 & wrongZ \\
SDSS J133338.30+041804.0 & 0.2022 & 18.52 & 18.26 & 17.70 & 17.11 & 17.10 & wrongZ\\
SDSS J133433.25$-$013825.4 & 0.2917 & 18.30 & 18.28 & 17.74 & 17.53 & 16.99 & NotGausSel, RL \\
SDSS J133957.99+613933.4 & 0.3723 & 19.04 & 18.79 & 18.63 & 18.60 & 17.92 & HiZ \\
SDSS J134617.55+622045.5 & 0.1163 & 16.92 & 16.83 & 16.57 & 16.02 & 16.06 & NotGausSel, NoParam, RL \\
SDSS J152139.66+033729.2 & 0.1261 & 16.94 & 16.91 & 16.74 & 15.98 & 16.31 & badFlag \\
SDSS J163545.62+481615.0 & 0.3088 & 19.09 & 19.23 & 18.89 & 18.86 & 18.12 & NotGausSel \\
SDSS J163856.54+433512.6 & 0.3391 & 18.44 & 18.39 & 18.22 & 18.19 & 17.44 & HiZ, RL \\
SDSS J171049.88+652102.2 & 0.3853 & 17.72 & 17.69 & 17.62 & 17.53 & 17.15 & HiZ \\
SDSS J205032.30$-$070131.2 & 0.1686 & 17.67 & 17.61 & 17.17 & 16.50 & 16.70 & NotGausSel \\
SDSS J212501.21$-$081328.6 & 0.6246 & 17.39 & 16.83 & 16.87 & 16.78 & 16.85 & HiZ, repeat1,22, NoParam, \ion{Mg}{2}Select \\
SDSS J215010.52$-$001000.7 & 0.3351 & 18.11 & 18.00 & 17.93 & 18.04 & 17.52 & HiZ \\
SDSS J222132.40$-$010928.7 & 0.2878 & 18.87 & 18.56 & 17.91 & 17.59 & 17.15 & badFlag \\
SDSS J223302.68$-$084349.1 & 0.0582 & 18.37 & 16.86 & 16.26 & 15.81 & 15.61 & NotPCASel \\
SDSS J230545.66$-$003608.6 & 0.2687 & 19.52 & 18.95 & 18.01 & 17.55 & 17.13 & NotGausSel, RL \\
SDSS J235128.76+155259.1 & 0.0966 & 17.55 & 16.85 & 16.24 & 15.72 & 15.63 & NotGausSel \\
\enddata
\label{tab2}\end{deluxetable}

\clearpage 
\begin{deluxetable}{ccccccccccc} 
  \tablewidth{0pt} \tabletypesize{\footnotesize} \tablenum{3}
  \tablecaption{H$\alpha$ profiles: Line Parameter Measurements}
  \tablehead{\colhead{SDSS ID} & \colhead{MJD} &\colhead{Flux}
    &\colhead{FWHM} &\colhead{FWHMc} &\colhead{FWQM} &\colhead{FWQMc}
    &\colhead{$H\si{red}$} &\colhead{$\lambda\si{red}$}
    &\colhead{$H\si{blue}$} &\colhead{$\lambda\si{blue}$}} \startdata
  SDSS J0007+0053 & 51456 & 7300 & 9800 & 400 & 12300 & 600 & 30.1 & 2300 & 33.9 & $-$2200 \\
  SDSS J0007+0053 & 51793 & 5500 & 9500 & 300 & 11900 & 400 & 22.5 & 2600 & 27.9 & $-$2200 \\
  SDSS J0008$-$1046 & 52141 & 1600 & 9600 & 1200 & 14700 & 0 & 6.7 & 2800 & 5.7 & $-$300 \\
  SDSS J0012$-$1022 & 52141 & 13400 & 5500 & 1200 & 9500 & 2100 & 41.2 & 4600 & 87.0 & 800 \\
  SDSS J0043+0051 & 51462 & 1900 & 11900 & 100 & 15200 & 500 & 5.8 & 2000 & 7.4 & $-$2600 \\
  SDSS J0043+0051 & 51794 & 4000 & 11800 & $-$200 & 14200 & $-$100 & 14.0 & 3600 & 18.8 & $-$3400 \\
  SDSS J0057+1446 & 51821 & 36200 & 9900 & 800 & 13300 & 1100 & 144.7 & 2600 & 144.9 & $-$1300 \\
  SDSS J0111$-$0958 & 52177 & 900 & 5400 & 100 & 7100 & 100 & 5.0 & 1400 & 6.7 & $-$900 \\
  SDSS J0117$-$0111 & 52202 & 3700 & 5600 & 200 & 7800 & 0 & 21.6 & 1500 & 26.8 & $-$700 \\
  SDSS J0132$-$0952 & 52147 & 3600 & 15300 & 200 & 18500 & 300 & 10.9 & 4600 & 10.7 & $-$3700 \\
  SDSS J0132$-$0952 & 52178 & 3900 & 15200 & 200 & 18400 & 100 & 12.2 & 4800 & 11.4 & $-$3500 \\
  SDSS J0134$-$0841 & 52147 & 1300 & 8300 & 200 & 11400 & 600 & 7.0 & 2000 & 5.7 & $-$1800 \\
  SDSS J0134$-$0841 & 52178 & 1300 & 8400 & 0 & 11700 & 500 & 7.2 & 1900 & 5.8 & $-$2000 \\
  SDSS J0149$-$0808 & 52174 & 1000 & 5400 & 0 & 7500 & 300 & 4.4 & 2200 & 8.4 & $-$600 \\
  SDSS J0216$-$0052 & 52209 & 1700 & 9600 & 100 & 11800 & 300 & 6.6 & 2600 & 8.6 & $-$600 
  \enddata   
  \tablecomments{This table is published in its entirety in the
    electronic edition of \emph{The Astrophysical Journal}. A portion
    is shown here for guidance regarding its form and content. ``MJD''
    in the second column refers to the last five digits of the
    Modified Julian Date. All positional measurements (FWHM, FWHMc,
    FWQM, FWQMc, the red peak position, $\lambda\si{red}$, and the
    blue peak position, $\lambda\si{blue}$) are in km\,s$^{-1}$
    relative to the narrow H$\alpha$ line; the line flux is in
    $10^{-17}$\,erg\,s$^{-1}$\,cm$^{-2}$, the peak heights,
    H$\si{red}$ and H$\si{blue}$, are in $10^{-17}$
    erg\,s$^{-1}$\,cm$^{-2}$\,\AA$^{-1}$.}
\label{tab3}\end{deluxetable}

\clearpage 
\begin{deluxetable}{cc} 
  \tablewidth{0pt} \tabletypesize{\footnotesize} \tablenum{4}
  \tablecaption{Errors in Measured Line Quantities} 
  \tablehead{\colhead{Parameter} &\colhead{Error}} 
  \startdata
  FWHM      &  6\%  \\
  FWHMc     &  200 km\,s$^{-1}$ \\
  FWQM      &  5\%  \\
  FWQMc     &  300 km\,s$^{-1}$ \\
  $H\si{red}$ and $H\si{blue}$ & 10\% \\
  $\lambda\si{red}$ & 30\% \\
  $\lambda\si{blue}$ & 40\% 
  \enddata 
   \tablecomments{Errors are estimated from repeat observations and
    processing as detailed in Section~\ref{sec:Merrors}. Errors in
    FWHM, FWQM, FWHMc, FWQMc, $\lambda\si{red}$ and $\lambda\si{blue}$ are
    measured in km\,s$^{-1}$ or percent thereof; peak height errors
    are in percent of flux density in units of $10^{-17}$
    erg\,s$^{-1}$\,cm$^{-2}$\,\AA$^{-1}$.}
\label{tab4}
\end{deluxetable}

\clearpage 
\begin{deluxetable}{lccccccccc} 
\tablewidth{0pt} \tabletypesize{\footnotesize} \tablenum{5}
\tablecaption{Optical, radio and X-ray luminosity and spectral
  indices} \tablehead{\colhead{SDSS ID} &
  \colhead{L\si{U}} &\colhead{L\si{G}} &\colhead{L\si{R}}
  &\colhead{L\si{I}} &\colhead{L\si{Z}} &\colhead{L\si{0.1-2\,keV}}
  &\colhead{L\si{20\,cm}}
  &\colhead{$\alpha\si{ox}$}&\colhead{$\alpha\si{or}$}} \startdata
SDSS J0007+0053 &      14.62 &    14.06 &    12.94 &     9.82 &    14.65 &    5.27 & \pt{ 6.72}{-5} &      1.4 &     0.10\\
SDSS J0008$-$1046 &     0.67 &     1.47 &     2.67 &     3.50 &     3.48 & \nodata & \nodata & \nodata & \nodata\\
SDSS J0012$-$1022 &     6.20 &     4.78 &     4.40 &     4.92 &     3.64 & \nodata & \pt{ 7.59}{-5} & \nodata &     0.14\\
SDSS J0043+0051 &       6.61 &     4.60 &     3.90 &     3.34 &     5.93 &    2.34 & \pt{ 7.06}{-5} &      1.5 &     0.11\\
SDSS J0057+1446 &       9.38 &     7.03 &     6.41 &     8.00 &     5.46 &    4.04 & \nodata &      1.5 & \nodata\\
SDSS J0111$-$0958 &     0.84 &     1.01 &     1.60 &     1.98 &     2.14 & \nodata & \nodata & \nodata & \nodata\\
SDSS J0117$-$0111 &     0.58 &     0.71 &     1.13 &     1.58 &     1.29 & \nodata & \nodata & \nodata & \nodata\\
SDSS J0132$-$0952 &     2.61 &     2.03 &     1.99 &     1.99 &     2.38 & \nodata & \nodata & \nodata & \nodata\\
SDSS J0134$-$0841 &     0.08 &     0.21 &     0.31 &     0.38 &     0.40 & \nodata & \nodata & \nodata & \nodata\\
SDSS J0149$-$0808 &     0.48 &     0.55 &     0.89 &     1.16 &     1.11 &    0.69 & \nodata &      1.2 & \nodata\\
SDSS J0212$-$0030 &    13.28 &    11.64 &     8.20 &     7.91 &     9.37 & \nodata & \pt{ 9.90}{-5} & \nodata &     0.12\\
SDSS J0216$-$0052 &     0.84 &     0.70 &     1.07 &     0.94 &     1.36 & \nodata & \nodata & \nodata & \nodata\\
SDSS J0220$-$0728 &     2.01 &     1.47 &     1.97 &     2.77 &     2.44 &    0.77 & \pt{ 2.70}{-5} &      1.5 &     0.14\\
SDSS J0229$-$0008 &     4.16 &     4.68 &     4.11 &     5.38 &     4.74 & \nodata & \pt{ 4.33}{-3} & \nodata &     0.54\\
SDSS J0232$-$0828 &     4.10 &     3.53 &     4.53 &     4.97 &     5.94 & \nodata & \nodata & \nodata & \nodata
\enddata 
\tablecomments{This table is published in its entirety in the
  electronic edition of \emph{The Astrophysical Journal}. A portion is
  shown here for guidance regarding its form and content. The
  luminosities are in units of $10^{44}$\,erg\,s$^{-1}$ computed using
  $\Omega_{\lambda}=0.73$, $\Omega\si{m}=0.27$, flat cosmology, with
  H\si{o}=72\,km\,s$^{-1}$\,Mpc$^{-1}$.}
\label{tab5}\end{deluxetable}

\clearpage 
\begin{deluxetable}{ccccccccc} 
  \tablewidth{0pt} \tabletypesize{\footnotesize} \tablenum{6}
  \tablecaption{Model Disks -- Grid Parameters}
  \tablehead{\colhead{Parameter} &\colhead{Value 1} &\colhead{Value 2}
    &\colhead{Value 3} &\colhead{Value 4} &\colhead{Value 5}
    &\colhead{Value 6} &\colhead{Value 7} &\colhead{Value 8}}
  \startdata
  $i$ [$^\circ$]    & 10 & 30  & 50 & 70  & 20 & ... & ... & ...  \\
  $q$      & $1.5$      & $2.0$       & $2.5$      & $3.0$       & ...  & ... & ... & ...  \\
  $\xi_1$ [$R\si{G}$] & $200$ & $300$  & $600$  & $800$  & ...  & ... & ... & ...  \\
  $\xi_2$ [$R\si{G}$] & $700$ & $1500$ & $2500$ & $4000$ & ...  & ... & ... & ...  \\
  $\sigma$ [km\,s$^{-1}$] & $780$ & $1200$ & $1800$ & $2400$  & ...  & ... & ... & ...  \\
  $e$      & $0$ & $0.2$  & $0.4$  & $0.7$ & ...  & ... & ... & ...  \\
  $\phi_0$ [$^\circ$] & $0$ & $45$ & $90$ & $140$ & $180$ & $230$ &
  $280$ & $320$ \enddata \tablecomments{$i$ is the inclination,
    with $i=90^\circ$ disks viewed edge on. The surface emissivity is
    parameterized by $q$, as $\epsilon_0\:\xi^{-q}$.  The inner and
    our radii, $\xi_1$ and $\xi_2$, are in units of the gravitational
    radius, $R\si{G}$=GM/c$^2$.  $\sigma$ is the local turbulent
    broadening, $e$ the ellipticity, and $\phi_0$ is the ellipse
    orientation, with $\phi_0=0^{\circ}$ equivalent to the apocenter
    pointing to the observer. $e=0$ refers to the circular disk models
    and the orientation angle $\phi_0$ is irrelevant. Also if
    $\xi_1=800$, then $\xi_2\neq700$.  This reduces the number of all
    possible arrangements of parameters from $4^6\times8=32,768$ to
    $24,000$.  The $i=20^\circ$ disk models were run only for circular
    disks ($\approx$ 960 additional models).}
 \label{tab6}\end{deluxetable}

\clearpage
\begin{deluxetable}{crrrrrrr} 
  \rotate \tablewidth{0pt} \tabletypesize{\footnotesize} \tablenum{7}
  \tablecaption{Inverse of the Covariance Matrix for the Observed Line
    Measurements} \tablehead{\colhead{} &\colhead{FWHM}
    &\colhead{FWQM} &\colhead{FWHMc} &\colhead{FWQMc}
    &\colhead{$\lambda\si{red}$} &\colhead{$\lambda\si{blue}$}
    &\colhead{H$_{Blue}/$$H\si{red}$}} \startdata
  FWHM & $2.4056\times10^{-6}$ & $-1.6648\times10^{-6}$ & $5.5465\times10^{-6}$ & $-2.8304\times10^{-6}$ & $-1.0930\times10^{-6}$ & $4.0709\times10^{-7}$ & $2.1411\times10^{-3}$ \\
  FWQM & $-1.6648\times10^{-6}$ & $1.5039\times10^{-6}$ & $-2.5392\times10^{-7}$ & $1.0518\times10^{-7}$ & $2.9943\times10^{-7}$ & $2.2996\times10^{-7}$  & $8.8773\times10^{-4}$ \\
  FWHMc & $5.5465\times10^{-7}$ & $-2.5392\times10^{-7}$ & $1.3838\times10^{-5}$ & $-8.3496\times10^{-6}$ & $-3.1662\times10^{-6}$ & $-2.0769\times10^{-6}$ & $1.0076\times10^{-2}$ \\
  FWQMc & $-2.8304\times10^{-7}$ & $1.0518\times10^{-7}$ & $-8.3496\times10^{-6}$ & $8.8003\times10^{-6}$ & $1.3741\times10^{-6}$ & $7.0378\times10^{-7}$ & $-6.5994\times10^{-3}$ \\
  $\lambda\si{red}$ & $-1.0930\times10^{-6}$ & $2.9943\times10^{-7}$ & $-3.1662\times10^{-6}$ & $1.3741\times10^{-6}$ & $2.2728\times10^{-6}$ & $-1.8431\times10^{-7}$ & $4.1504\times10^{-3}$ \\
  $\lambda\si{blue}$ & $4.0709\times10^{-7}$ & $2.2996\times10^{-7}$ & $-2.0769\times10^{-6}$ & $7.0378\times10^{-7}$ & $-1.8431\times10^{-7}$ & $2.0338\times10^{-6}$ & $-1.1646\times10^{-3}$ \\
  H$_{Blue}/$$H\si{red}$ & $2.1411\times10^{-2}$ &
  $-8.8773\times10^{-4}$ & $1.0076\times10^{-2}$ &
  $-6.5994\times10^{-3}$ & $-4.1504\times10^{-3}$ &
  $-1.1646\times10^{-3}$ & $1.5947\times10^{1}$ \enddata
  \tablecomments{The values quoted here assume FWHM, FWHMc, FWQM,
    FWQMc, $\lambda\si{red}$, and $\lambda\si{blue}$ measured in
    km\,s$^{-1}$ with respect to the narrow line H$\alpha$ and peak
    height ratio, H$_{Blue}/$$H\si{red}$, with peak heights measured
    in erg\,s$^{-1}$\,cm$^{-2}$\,\AA$^{-1}$.}
 \label{tab7}\end{deluxetable}

\clearpage
\begin{deluxetable}{cll}
  \tablewidth{0pt} \tabletypesize{\footnotesize} \tablenum{8}
  \tablecaption{Vacuum Wavelengths}
  \tablehead{\colhead{Line} &\colhead{Air} &\colhead{Vacuum}}
  \startdata
  \mbox{[\ion{O}{3}]} & 4958.91\AA & 4960.30\AA \\
  H$\beta$   & 4861.36\AA & 4862.72\AA \\
  \mbox{[\ion{O}{3}]} & 5006.84\AA & 5008.24\AA \\
  \mbox{[\ion{O}{1}]} & 6302.05\AA & 6300.30\AA \\
  \mbox{[\ion{O}{1}]} & 6365.54\AA & 6363.78\AA \\
  \mbox{[\ion{N}{2}]} & 6548.05\AA & 6549.86\AA \\
  H$\alpha$  & 6562.80\AA & 6564.61\AA \\
  \mbox{[\ion{N}{2}]} & 6583.45\AA & 6585.27\AA \\
  \mbox{[\ion{S}{2}]} & 6716.44\AA & 6718.29\AA \\
  \mbox{[\ion{S}{2}]} & 6730.82\AA & 6732.68\AA
  \enddata
\label{tab8}\end{deluxetable}

\end{document}